\documentclass[letterpaper,11pt]{article}
\pdfoutput=1
\usepackage{jcappub}
\usepackage{etoolbox}
 \makeatletter
    \patchcmd{\maketitle}{\@fpheader}{}{}{}
    \makeatother
\usepackage{threeparttable} 
\usepackage{color}
\usepackage{epstopdf}
\usepackage{natbib}
\usepackage{mathrsfs}
\usepackage{subcaption}
\usepackage{tikz-cd}
\usepackage{mdwlist}
\usepackage{ulem}

\usepackage{dsfont}
\usepackage{natbib}
\usepackage{amssymb}
\usepackage{amsthm}
\usepackage{mathtools}
\usepackage{graphicx}
\usepackage{amsmath}
\usepackage{url}
\usepackage{bm}
\usepackage{epstopdf}
\usepackage{color}
\usepackage{enumitem}
\usepackage{dsfont}
\usepackage{bbm}
\usepackage{mathrsfs}
\usepackage[margin=1.in]{geometry}
\usepackage[utf8]{inputenc}
\usepackage{url}
\usepackage{stackrel}
\usepackage{accents}

\newcommand\munderbar[1]{%
  \underaccent{\bar}{#1}}

\newcommand{\rnew}{r_{\scalebox{0.65}{0}}}

\newcommand{\Tr}{\mbox{Tr}}
\newcommand{\be}{\begin{equation}}
\newcommand{\ee}{\end{equation}}
\newcommand{\de}{\mbox{d}}

\newcommand{\pa}{\partial}

\newcommand{\scr}{\scriptscriptstyle}

\newcommand{\cmmnt}[1]{}

\numberwithin{equation}{section}

\begin{document}

\title{Gravitational wave oscillations in bimetric cosmology}

\author[a]{David Brizuela,}
\author[b,c]{Marco de Cesare,}
\author[a]{Araceli Soler Oficial}

\emailAdd{david.brizuela@ehu.eus}
\emailAdd{marco.decesare@na.infn.it}
\emailAdd{araceli.soler@ehu.eus}

\affiliation[a]{Department of Physics and EHU Quantum Center, University of the Basque Country UPV/EHU, Barrio Sarriena s/n, 48940 Leioa, Spain}

\affiliation[b]{Dipartimento di Fisica ``Ettore Pancini'', Universit\`a di Napoli Federico II, Napoli, Italy}

\affiliation[c]{INFN, Sezione di Napoli, Italy}

\abstract{Unlike general relativity, in bimetric gravity linear gravitational 
waves do not evolve as free fields. In this theory there are two types 
of tensor perturbations, whose interactions are inherited from 
non-trivial couplings between two dynamical metric tensor fields in 
the Hassan-Rosen action, and are responsible for the phenomenon of 
{\it bigravity oscillations}.
In this work, we analyze the dynamics of cosmological tensor modes in 
bimetric gravity on sub-horizon scales and close to the {\it general 
relativity limit}.
In this limit, the system has a characteristic length scale $L$ that is strictly contained within the comoving Hubble radius.
Thus, depending on the magnitude of the 
comoving wavelength $\lambda$ relative to $L$, 
we identify two regimes of interest where the system can be studied analytically: (i) deep 
sub-horizon modes with $\lambda\ll L$,
whose dynamics can be studied using multiple scale analysis and are characterized by small and slowly evolving
super-imposed perturbations;
(ii) sub-horizon modes with $\lambda\gg L$, 
where the dynamics is characterized by fast super-imposed oscillations 
that can be studied using asymptotic techniques for highly oscillatory 
problems.
Furthermore, our analysis represents a substantial improvement 
compared to previous analyses based on a generalization of the WKB 
method, which, as we show, is ill-suited to study the system at hand.
}

\maketitle

\nopagebreak

\section{Introduction}

In the context of general relativity (GR) the gravitational interaction is mediated by the graviton, a massless spin-2 field with non-linear self-interactions~\cite{Deser:1969wk}. Despite the remarkable experimental and observational evidence in favour of GR~\cite{WILL:2014kxa,Yunes:2013dva,LIGOScientific:2020tif} and the success of the $\Lambda$CDM cosmological model \cite{Planck:2018vyg}, the status of GR as a fundamental theory of gravity has been challenged by modified gravity theories \cite{Clifton:2011jh,Berti:2015itd}. Such theories typically lead to modifications in the behaviour of gravity on large scales that could provide an alternative to dark energy and dark matter.
Moreover, it has been argued that modified gravity may alleviate or even solve cosmological tensions recently emerged in the $\Lambda$CDM model~\cite{DiValentino:2021izs}.

Bimetric gravity is a highly non-trivial modification of GR that describes consistent non-linear interactions between massive and massless spin-2 fields \cite{Hassan:2011vm, Hassan:2011tf} (see also Ref.~\cite{Schmidt-May:2015vnx}). The theory features two dynamical metric fields with suitable interaction terms that ensure the absence of a Boulware-Deser ghost, which plagued previous attempts to formulate a consistent non-linear bimetric theory~\cite{Boulware:1972yco}. In the limit where one of the two metrics becomes non-dynamical, one recovers de Rham, Gabadadze and Tolley (dRGT) massive gravity \cite{deRham:2010kj}. The theory admits yet another independent limit, whereby GR is recovered \cite{Akrami:2015qga}. It has been shown that bimetric gravity is stable and well behaved in certain regions of parameter space \cite{Fasiello:2013woa}, and that viable cosmological solutions that fit the expansion history of the accelerating universe also exist \cite{Volkov:2011an, Volkov:2012zb, DeFelice:2014nja, Akrami:2015qga}. Furthermore, bimetric theory has been confronted with observations of Type Ia Supernovae, Baryon Acoustic Oscillations and the Cosmic Microwave Background \cite{vonStrauss:2011mq, Caravano:2021aum, Hogas:2021lns}, which constrain the parameters of the theory. Further constraints can be obtained from Big Bang nucleosynthesis \cite{Hogas:2021saw} and from the requirement that there is a working screening mechanism \cite{Hogas:2021fmr}. 

The focus of this work is on the propagation of gravitational waves in an expanding universe in bimetric gravity, which leads to effects that are potentially testable with LIGO/Virgo gravitational wave observatories and with the forthcoming LISA mission, which will provide data in a yet unexplored frequency band~\cite{LISA:2017pwj}. At a perturbative level, the bimetric structure of the theory leads to two distinct types of tensor perturbations (one for each metric) that couple to each other, with time-dependent couplings determined by the cosmological background. In a de~Sitter universe, the couplings become constant and the evolution of tensor modes is characterized by an oscillatory behaviour analogous to neutrino oscillations, dubbed {\it gravitational wave oscillations} \cite{Max:2017flc, Max:2017kdc}.
Although cosmological perturbations and their propagation have been previously studied by different groups \cite{Comelli:2012db, DeFelice:2013nba, Lagos:2014lca, Amendola:2015tua, Cusin:2014psa}, general analytical methods are still needed to solve the dynamics for general background evolution and for generic values of the free parameters of the model. A proposal to generalize the standard WKB method to the case of interacting oscillators was made in Ref.~\cite{BeltranJimenez:2019xxx}, and applications to bimetric gravity have been studied in Refs.~\cite{BeltranJimenez:2019xxx,LISACosmologyWorkingGroup:2019mwx,Ezquiaga:2021ler}. However, unlike the standard WKB method, the regimes of applicability of such a generalization remain unclear, due to subtle issues that have been overlooked in previous analyses (these are discussed in detail in Appendix~\ref{Sec:AppendixWKB}).

In this work we propose a different method. We focus specifically on the evolution of sub-horizon tensor modes on a background dominated by hydrodynamic matter and in the GR limit (as defined in Ref.~\cite{Akrami:2015qga}). Moreover, to keep our analysis as general as possible, we do not assume any specific values for the parameters of the theory in the derivation of approximate analytical solutions. Next, we identify two main regimes of interest for the dynamics of sub-horizon modes, that are separated by a critical scale $L=\alpha(a H)^{-1}$ determined by the product of the comoving Hubble radius and the parameter $\alpha$ that governs the GR limit. We show that deep sub-horizon modes with comoving wavelength $\lambda\ll\alpha(a H)^{-1}$ can be studied using multiple-scale analysis \cite{kevorkian1996multiple}, which is well suited for perturbative problems with dependence on very different evolution timescales and gives far more reliable solutions compared to the generalized WKB method used in Refs.~\cite{BeltranJimenez:2019xxx,LISACosmologyWorkingGroup:2019mwx,Ezquiaga:2021ler}. Moreover, we develop a novel method to study intermediate scales, i.e.~for modes with $\alpha(a H)^{-1}\ll\lambda\ll(a H)^{-1}$, which are characterized by fast superimposed oscillations.

The remainder of this paper is organized as follows. In Section~ \ref{Sec: BimetricReview} we review the formulation of bimetric gravity and its GR limit. The dynamical equations for the cosmological background and tensor perturbations are reviewed in Section~\ref{Sec: TensorPerturbations}. Significant simplifications in the structure of the equations are obtained in Section~\ref{Sec: DynamicsSubHorizon}, where we exploit the properties of the background evolution on the so-called {\it finite branch}. Our original results are contained in Section~\ref{Sec: AnalyticalSolutions} where we bring the system to a suitable form for the application of multiple-scale analysis. Here we identify the two regimes of interest and characterize them physically. After bringing the system of coupled oscillators to action-angle variables, we solve it using suitable analytical techniques in each regime. Our results are then compared with numerical simulations, showing that the agreement is excellent. Lastly, we review our results in Section~\ref{Sec: Conclusion}.

Technical appendices are also included. In particular, Appendix~\ref{Sec:AppendixSecondOrderAction} contains a derivation of the second order action for tensor perturbations in a curved background; although this result has been derived previously in Ref.~\cite{DeFelice:2014nja}, here we illustrate the calculations in full detail. Lastly, Appendix~\ref{Sec:AppendixWKB} includes a brief critical review of the generalized WKB method of Refs.~\cite{BeltranJimenez:2019xxx,LISACosmologyWorkingGroup:2019mwx,Ezquiaga:2021ler} and a detailed comparison with the multiple scale solution for a toy model with the same structure of the coupled oscillators in bigravity.

\vspace{0.5cm}

\noindent {\bf Notations and conventions:}  We assume the metric signature $(-+++)$ and units with $c=1$.
Greek indices are used for spacetime tensors, while Latin indices are used for spatial tensors.

\section{Bimetric gravity: action and field equations}\label{Sec: BimetricReview}

The Hassan-Rosen bimetric gravity action reads \cite{Hassan:2011zd}
\be\label{Eq:BimetricAction}
S_{\rm\scriptscriptstyle HR}[g_{ab},f_{ab}]=\frac{M_g^2}{2} \int \de^4 x \sqrt{-g}\, R^{(g)}+\frac{M_f^2}{2} \int \de^4 x \sqrt{-f}\, R^{(f)}-m^2 M_g^2 \int \de^4x \sqrt{-g}\,\sum_{n=0}^4 \beta_n e_n(\mathbb{S})~,
\ee
where $R^{(g)}$ and $R^{(f)}$ are the Ricci scalars of the metrics $g_{ab}$ and $f_{ab}$, respectively. Similarly, in the following, we will use the superscripts $^{(g)}$ and $^{(f)}$ to distinguish elements corresponding to each metric.
The constants $M_g$, $M_f$, and $m$ have physical dimensions of mass, whereas the constants $\beta_n$ are dimensionless, and the matrix $\mathbb{S}$ is defined in terms of the two metrics $g_{ab}$ and $f_{ab}$ as
\be
\mathbb{S}^a_{\;b}=\sqrt{g^{ac}f_{cb}}~.
\ee
Finally, the symmetric polynomials of $\mathbb{S}$ are defined as \cite{Hassan:2011hr,Bernard:2015mkk}
\begin{subequations}\label{Eq:SymmetricPoly_Defs}
\begin{align}
e_0 (\mathbb{S})&=1~,\\
e_1 (\mathbb{S})&=\Tr[\mathbb{S}] ~,\\
e_2 (\mathbb{S})&=\frac{1}{2}\left(\Tr[\mathbb{S}]^2 -  \Tr[\mathbb{S}^2]\right)~,\\
e_3 (\mathbb{S})&=\frac{1}{6}\left(\Tr[\mathbb{S}]^3 - 3\Tr[\mathbb{S}] \Tr[\mathbb{S}^2]+2\Tr[\mathbb{S}^3]\right)~,\\
e_4 (\mathbb{S})&=\frac{1}{24}\left(\Tr[\mathbb{S}]^4 - 6\Tr[\mathbb{S}]^2 \Tr[\mathbb{S}^2]+3\Tr[\mathbb{S}^2]^2 + 8 \Tr[\mathbb{S}] \Tr[\mathbb{S}]^3 - 6\Tr[\mathbb{S}^4]  \right)~,
\end{align}
\end{subequations}
with $\Tr[\mathbb{S}]=\mathbb{S}^a_{\;a}$.

\

In this framework, matter fields are usually assumed to couple only to the metric $g_{ab}$, which ensures the absence of the Boulware-Deser ghost\footnote{Note however that, in the low-energy effective field theory \cite{deRham:2014naa} it is possible to find a combination of both metrics for which the ghost is not excited and matter could be coupled to both metrics.} \cite{Yamashita:2014fga,deRham:2014naa}.
Thus, the total action reads
\be\label{Eq:Action_gravity}
S[g_{ab},f_{ab},\psi]=S_{\rm\scriptscriptstyle HR}[g_{ab},f_{ab}]+S_{\rm\scriptscriptstyle m}[g_{ab},\psi]~,
\ee
where matter fields are collectively denoted as $\psi$, and are minimally coupled to gravity.
The field equations obtained from the variation of Eq.~\eqref{Eq:Action_gravity} read
\begin{subequations}\label{Eq:BigravityEom_matter}
\begin{align}\label{Eq:BigravityEom_matter-g}
R_{\mu\nu}^{(g)}-\frac{1}{2}g_{\mu\nu}R^{(g)}+m^2V_{\mu\nu}^{(g)}(g,f,\beta_n)&=\frac{1}{M_g^2}T_{\mu\nu}~,\\\label{Eq:BigravityEom_matter-f}
R_{\mu\nu}^{(f)}-\frac{1}{2}f_{\mu\nu}R^{(f)}+\frac{m^2}{\alpha^2}V_{\mu\nu}^{(f)}(g,f,\beta_n) &=0~,
\end{align}
\end{subequations}
where $\alpha\equiv M_f/M_g$ is the ratio between the `Planck masses' of the two metrics,
and $T_{\mu\nu}$ is the matter stress-energy tensor, defined as usual by
\be
T_{\mu\nu}\equiv -\frac{2}{\sqrt{-g}}\frac{\delta S_{\rm\scriptscriptstyle m}}{\delta g^{\mu\nu}}~.
\ee
The interaction potentials are given by
\begin{subequations}\label{Eq:InteranctionPotentials}
\begin{align}
    V_{\mu\nu}^{(g)}(g,f,\beta_n)&=\frac{1}{2}\sum_{i=0}^3(-1)^i\beta_i\left[g_{\mu\rho}(Y_{(i)})^{\rho}_{\,\,\nu}(\mathbb{S})+g_{\nu\rho}(Y_{(i)})^{\rho}_{\,\,\mu}(\mathbb{S})\right]~,\\
    V_{\mu\nu}^{(f)}(g,f,\beta_n)&=\frac{1}{2}\sum_{i=0}^3(-1)^i\beta_{4-i}\left[f_{\mu\rho}(Y_{(i)})^{\rho}_{\,\,\nu}(\mathbb{S}^{-1})+f_{\nu\rho}(Y_{(i)})^{\rho}_{\,\,\mu}(\mathbb{S}^{-1})\right]~,
\end{align}
\end{subequations}
where the matrices $Y_{(i)}$ read
 \begin{equation}\label{Eq:YMatrices}
    (Y_{(i)})^{\rho}_{\,\,\nu}(\mathbb{S})\equiv \sum_{k=0}^i(-1)^k(\mathbb{S}^{i-k})^{\rho}_{\,\,\nu}e_k(\mathbb{S})~.
\end{equation}

The matter action $S_{\rm\scriptscriptstyle m}$ is assumed to be diffeomorphism invariant, which ensures the conservation of energy-momentum $\nabla^{\mu}T_{\mu\nu}=0$. As a consequence of the Bianchi identity $g^{\mu\rho}\nabla_{\rho} G_{\mu\nu}^{(g)}=0$,
the divergence of \eqref{Eq:BigravityEom_matter-g} gives the Bianchi constraint
\be\label{Eq:BianchiConstraintV}
g^{\mu\rho}\nabla_{\rho} V_{\mu\nu}^{(g)}=0~,
\ee
where $\nabla$ is the covariant derivative compatible with $g_{\mu\nu}$. The corresponding relations for $f_{\mu\nu}$ do not provide additional information, given that the divergences of the interaction contributions satisfy~\cite{Damour:2002ws, Schmidt-May:2015vnx} 
\begin{equation}\label{Eq:BianchiConstraintVtilde}
    \sqrt{-g}g^{\mu\rho}\nabla_{\rho}V^{(g)}_{\mu\nu}=-\sqrt{-f}f^{\mu\rho}\widetilde{\nabla}_{\rho}V^{(f)}_{\mu\nu}~.
\end{equation}
Here $\widetilde{\nabla}$ is the covariant derivative compatible with $f_{\mu\nu}$. Thus, Eq.~\eqref{Eq:BianchiConstraintV} implies that also the right-hand side of Eq.~\eqref{Eq:BianchiConstraintVtilde} must vanish.

\subsection{The general-relativity limit}\label{Sec:GRlimit}
In this section we review the argument given in Ref.~\cite{Hassan:2014vja, Akrami:2015qga} that shows how general relativity can be recovered from bimetric gravity in the limit where $\alpha$ tends to zero.
First, note that the bimetric interaction potentials \eqref{Eq:InteranctionPotentials} satisfy the following identity:
\begin{equation}\label{Eq:identityGRlimit}
    \sqrt{-g}g^{\mu\rho}V^{(g)}_{\,\rho\nu}+\sqrt{-f}f^{\mu\rho}V^{(f)}_{\,\rho\nu}-\sqrt{-g}V\delta^{\mu}_{\,\,\,\nu}=0\,,
\end{equation}
where $V$ is the interaction term in the bimetric action \eqref{Eq:BimetricAction}, that is,
\begin{equation}\label{Eq:DefV}
    V=\sum_{n=0}^{4}\beta_ne_n\left(\mathbb{S}\right)\,.
\end{equation}
Plugging the field equations \eqref{Eq:BigravityEom_matter} into the relation \eqref{Eq:identityGRlimit}, we obtain
\begin{equation}
    g^{\mu\rho}\left(R_{\rho\nu}^{(g)}-\frac{1}{2}g_{\rho\nu}R^{(g)}\right)+\alpha^2\sqrt{g^{-1}f}f^{\mu\rho}\left(R_{\rho\nu}^{(f)}-\frac{1}{2}g_{\rho\nu}R^{(f)}\right)+m^2V\delta^{\mu}_{\nu}=\frac{1}{M_g^2}g^{\mu\rho}T_{\,\rho\nu}\,,
\end{equation}
which, in the limit $\alpha\to 0$, reduces to
\begin{equation}\label{Eq:GRLimit-g}
    R_{\mu\nu}^{(g)}-\frac{1}{2}g_{\mu\nu}R^{(g)}+m^2Vg_{\mu\nu}=\frac{1}{M_g^2}T_{\,\mu\nu}\,.
\end{equation}
Next, taking the covariant divergence of \eqref{Eq:GRLimit-g}, as the Einstein tensor and the source $T_{\,\mu\nu}$ are both covariantly conserved, we deduce that  $\nabla_{\mu} V{\big |}_{\alpha\to 0}=0$.
Therefore, in this limit $V$ is constant on-shell, and Eq.~\eqref{Eq:GRLimit-g} can thus be interpreted as Einstein equations for the metric $g_{\mu\nu}$ with a non-zero cosmological constant $m^2V$. From here, we conclude that the general-relativity limit of bimetric gravity is defined by $\alpha\to 0$, and that bigravity's self-acceleration survives in this limit---even for $\beta_0=0$ \cite{Akrami:2015qga}.
As can be seen in \eqref{Eq:BigravityEom_matter-f}, this is a strong-coupling limit for the $f_{\mu\nu}$ sector,
which is completely determined in terms of $g_{\mu\nu}$ \cite{Hassan:2014vja}.

Let us end this section by commenting on another limit of the theory: $m\to 0$. Looking at the form of the Hassan-Rosen action \eqref{Eq:BimetricAction},
it is clear that in the weak-coupling limit $m\to 0$, each metric $g_{\mu\nu}$ and $f_{\mu\nu}$ evolves independently
following its corresponding Einstein equations. However, this limit suffers from the so-called van Dam-Veltman-Zakharov discontinuity \cite{vanDam:1970vg,Zakharov:1970cc}. On the contrary, the limit $\alpha\to 0$ is not affected by such
discontinuity.

\section{Linear tensor perturbations of cosmological solutions in bimetric gravity}\label{Sec: TensorPerturbations}

This section is divided in two subsections. In Sec.~\ref{Sec:Background}
we consider a homogeneous and isotropic solution of the equations of motion of bimetric
gravity \eqref{Eq:BigravityEom_matter}, with the matter content given by a perfect fluid.
Subsequently, in Sec.~\ref{Sec:perturbations}, we analyze the dynamics of tensor perturbations
evolving on the commented cosmological background.

\subsection{Dynamics of the cosmological background}\label{Sec:Background}

Let us assume the closed Friedman-Lemaître-Robertson-Walker (FLRW) geometry as a background for the two metrics $f_{ab}$ and $g_{ab}$. As shown in Ref.~\cite{Nersisyan:2015oha}, the Bianchi constraints imply that the curvature radii for the spatial geometries must be the same for the two background metrics. Thus, we can write
\begin{subequations}\label{Eq:BackgroundFLRWmetrics}
\begin{align}
\de s_{(g)}^2&=a^2(\eta) \left[ -\de \eta^2 +\gamma_{ij}\de x^i \de x^j \right]~,\\
\de s_{(f)}^2&=b^2(\eta) \left[ -c(\eta)^2 \de \eta^2 +\gamma_{ij}\de x^i \de x^j \right]~.
\end{align}
\end{subequations}
Here $a$ and $b$ are the scale factors, $c$ is the lapse of the conformal time of $f_{ab}$ relative to $g_{ab}$, and $\gamma_{ij}$ is the metric on the three-sphere $\gamma_{ij}\,\text{d} x^i \text{d} x^j=\rnew^2\left[\text{d}\chi^2+\sin^2{\chi}\left(\text{d}\theta^2+\sin^2{\theta}\,\text{d}\phi^2\right)\right]$, with a fixed radius $\rnew$. The conformal Hubble rates of $g_{ab}$ and $f_{ab}$ will be denoted as $\mathcal{H}\equiv a^{\prime}/a$ and $\mathcal{H}_f \equiv b^{\prime}/(b c)$, respectively,
where the prime stands for a derivative with respect to the conformal time $\eta$. 

The equations of motion for the two metrics, obtained from \eqref{Eq:BigravityEom_matter}, read
\begin{subequations}\label{Eq:BackgroundEqs}
\begin{align}
&\mathcal{H}^2+\frac{1}{\rnew^2}=\frac{1}{3 }M_g^{-2}a^2(\rho_m+\rho_g)~, \label{Eq:BackgroundEqs_g-1}\\
& \mathcal{H}^2+2\mathcal{H}^{\prime}+ \frac{1}{\rnew^2}=-M_g^{-2} a^2 (p_m+p_g) ~,\label{Eq:BackgroundEqs_g-2}\\
&\mathcal{H}_f^2+\frac{1}{\rnew^2}=\frac{1}{3 }M_f^{-2}b^2\rho_f~,\label{Eq:BackgroundEqs_f-1}\\
&\mathcal{H}_f^2+\frac{2}{c}\mathcal{H}_f^{\prime}+ \frac{1}{\rnew^2}=-M_f^{-2} b^2 p_f ~\label{Eq:BackgroundEqs_f-2},
\end{align}
\end{subequations}
where $\rho_m$ and $p_m$ are the energy density and pressure of the perfect fluid, respectively, and satisfy the continuity equation 
\begin{equation}\label{Eq:ContinuityEquationMatter}
\rho_m^{\prime}+3\mathcal{H}(\rho_m+p_m)=0~.
\end{equation}
Deviations from general relativity due to bigravity interactions are encoded in the effective energy densities and pressures, which are defined as
\begin{subequations}\label{Eq:EffectiveFluids}
\begin{align}
\rho_g&= m^2 M_g^2 \left(\beta_0+3 \beta_1 y+3 \beta_2 y^2+\beta_3 y^3 \right)~,\\
 p_g&= -m^2 M_g^2 \left(\beta_0+ \beta_1 (2+c) y+ \beta_2 (1+2 c) y^2+\beta_3 c\, y^3 \right)~,\\
\rho_f&= m^2 M_g^2 \left(\beta_1 y^{-3}+3 \beta_2 y^{-2}+3 \beta_3 y^{-1}+\beta_4 \right)~,\\
 p_f&= -m^2 M_g^2 \left(\beta_1 c^{-1} y^{-3}+ \beta_2 (1+2 c^{-1}) y^{-2}+ \beta_3 (2+c^{-1})y^{-1}+\beta_4  \right)~,
\end{align}
\end{subequations}
with $y\equiv b/a$ representing the ratio between the scale factors. We assume that matter obeys the null energy condition as a strict inequality, $\rho_m+p_m>0$.

The Bianchi constraint \eqref{Eq:BianchiConstraintV} is equivalent to the continuity equation for the effective fluid, $\rho_g^{\prime}+3\mathcal{H}(\rho_g+p_g)=0$, which gives
\be\label{Eq:BianchiConstraint}
\Gamma(y)(\mathcal{H}_f-\mathcal{H})=0 ~,
\ee
where $\Gamma(y)\equiv \beta_1+2\beta_2 y+\beta_3 y^2$.
As it is clear written in the form ~\eqref{Eq:BianchiConstraint},
there are two possible solutions to the Bianchi constraint. On the one hand, it is satisfied if the conformal Hubble rates for the two metrics coincide, $\mathcal{H}_f=\mathcal{H}$: this condition defines the so-called {\it dynamical branch}. On the other hand, \eqref{Eq:BianchiConstraint} is also satisfied if $\Gamma(y)=0$, which defines the so-called {\it algebraic branch} \cite{Schmidt-May:2015vnx}. However, it is well known that solutions on this latter branch are unstable \cite{DeFelice:2014nja,Cusin:2015tmf}. Hence, we will not take them into account and, in the remaining of the paper,
we will focus on the dynamical branch with $\mathcal{H}_f=\mathcal{H}$.

Therefore, using the the condition ${\cal H}_f={\cal H}$, in combination with the two effective Friedmann equations in~\eqref{Eq:BackgroundEqs}, it is possible to obtain an algebraic relation
between the different energy densities,
\be\label{Eq:Densities_Relation}
\rho_m+\rho_g = \frac{y^2}{\alpha^2}  \rho_f ~.
\ee
Replacing in this relation the definitions of the effective energy densities \eqref{Eq:EffectiveFluids}
leads to the following polynomial equation for the ratio between the scale factors $y$
\be\label{Eq:yAlgebraic}
\frac{\beta_1}{\alpha^2} + \left(\frac{3\beta_2}{\alpha^2} - \beta_0-\frac{\rho_m}{m^2 M_g^2}\right)y+3\left(\frac{\beta_3}{\alpha^2}  -\beta_1\right)y^2+\left( \frac{\beta_4}{\alpha^2} -3 \beta_2\right)y^3-\beta_3\, y^4=0~.
\ee

An evolution equation for the ratio $y$ can also be obtained from the definitions of the conformal Hubble rates for the two metrics
\be\label{Eq:BianchiConstraintRewrite0}
\frac{y^{\prime}}{y}=\left(c \mathcal{H}_f-\mathcal{H} \right)~.
\ee
Using the Bianchi constraint on the dynamical branch, ${\cal H}_f={\cal H}$,
this equation simplifies to
\be\label{Eq:BianchiConstraintRewrite1}
\frac{y^{\prime}}{y}=(c-1) \mathcal{H} ~,
\ee
which can equivalently be written as
\be\label{Eq:BianchiConstraintRewrite2}
\frac{ \de (\log y)}{\de (\log a)} = c-1~.
\ee

\subsection{Dynamics of tensor perturbations}\label{Sec:perturbations}

As in general relativity, also in the theory of bimetric gravity, scalar, vector and tensor perturbations evolve independently at the level of the linearized field equations.
In this work we focus exclusively on tensor perturbations evolving on the background
described in the previous section. Thus, our perturbative ansatz
can be written as
\begin{subequations}\label{Eq:PerturbedFLRWmetrics}
\begin{align}
\de s_{(g)}^2&=a^2(\eta) \left[ -\de \eta^2 +\left(\gamma_{ij}+h_{ij}(\eta,x)\right)\de x^i \de x^j \right]~,\\
\de s_{(f)}^2&=b^2(\eta) \left[ -c(\eta)^2 \de \eta^2 +\left(\gamma_{ij}+H_{ij}(\eta,x)\right)\de x^i \de x^j \right]~,
\end{align}
\end{subequations}
where $h_{ij}$ and $H_{ij}$ are tensor perturbations, i.e.,~they are transverse and traceless
\begin{align}
&\gamma^{ij}h_{i j}=\gamma^{ij}H_{i j}=0~,\\ &\gamma^{jk}\mathcal{D}_k h_{ij}=\gamma^{jk}\mathcal{D}_k H_{ij}=0~.
\end{align}
Here $\mathcal{D}_i$ denotes the Levi-Civita connection associated with $\gamma_{ij}$, $\mathcal{D}_i \gamma_{jk}=0$. In the following, spatial indices will be raised and lowered with the metric $\gamma_{ij}$.

The equations of motion can be obtained by expanding the field equations~\eqref{Eq:BigravityEom_matter} to linear order in $h_{ij}$ and $H_{ij}$, using the perturbative ansatz \eqref{Eq:PerturbedFLRWmetrics},
\begin{subequations}\label{Eq:TensorPert}
\begin{align}
&h^{\prime\prime}_{ij} +2 \mathcal{H}\, h^{\prime}_{ij}- \mathcal{D}_k\mathcal{D}^k h_{ij} + \frac{2}{\rnew^2}h_{ij}+m^2 a^2 \lambda(y) (h_{ij}-H_{ij}) =\frac{2}{M_g^2} a^2 \pi_{ij}~,\label{Eq:TensorPert1} \\
&H^{\prime\prime}_{ij} +\left(2 c\, \mathcal{H}_f -\frac{c^{\prime}}{c}\right) H^{\prime}_{ij} - c^2 \mathcal{D}_k\mathcal{D}^k H_{ij} + \frac{2 c^2}{\rnew^2}H_{ij}+\frac{m^2}{\alpha^2} b^2 c\, y^{-4}\lambda(y) (H_{ij}-h_{ij}) =0~,\label{Eq:TensorPert2} 
\end{align}
\end{subequations}
where we have defined the third-order polynomial in the ratio $y$,
\be\label{Eq:LambdaDefinitions}
\lambda(y)\equiv \beta_1 y+\beta_2 (1+c)y^2+\beta_3 c\, y^3 ~,
\ee
and $\pi_{ij}$ is the anisotropic stress, i.e., the tensorial component of the linear perturbations of $T_{\mu\nu}$.

\subsubsection*{Harmonic decomposition on $S^3$}

In order to simplify the above equations, and taking into account the symmetries of the background, we decompose the perturbations in tensor spherical harmonics \cite{Gerlach:1978gy},
\begin{subequations}\label{Eq:GSmetricperts}
\begin{align}
h_{ij}(\eta,\chi,\theta,\phi)=\sum_{s=0}^{1} \sum_{n=3}^{+\infty} \sum_{l=2}^{n-1} \sum_{m=-l}^{l}  h_{n l m s}(\eta) Q^{n l m s}_{ij}(\chi,\theta,\phi) ~,\\
H_{ij}(\eta,\chi,\theta,\phi)=\sum_{s=0}^{1} \sum_{n=3}^{+\infty} \sum_{l=2}^{n-1} \sum_{m=-l}^{l}  H_{n l m s}(\eta) Q^{n l m s}_{ij}(\chi,\theta,\phi) ~.
\end{align}
\end{subequations}
The index $s$ represents the polarity ($s=0$ for even/polar harmonics and $s=1$ for odd/axial harmonics),
and the $Q^{n l m s}_{ij}$ are eigenfunctions of the Laplacian on the three-sphere with radius $\rnew$,
\be
\mathcal{D}_k\mathcal{D}^k Q^{n l m s}_{ij}= - \frac{(n^2-3)}{\rnew^2} Q^{n l m s}_{ij} ~.
\ee
By definition, these are transverse and traceless: $\mathcal{D}^i Q^{n l m s}_{ij}=0$, $\gamma^{ij}Q^{n l m s}_{ij}=0$. Note that there are two main differences with tensor modes in a flat universe: (i) the spectrum of the spatial Laplacian is discrete rather than continuous; (ii) the minimum eigenvalue is strictly positive \cite{Challinor:2000as}.

A similar decomposition is introduced for the anisotropic stress-energy tensor
\be\label{Eq:GSmatterpert}
\pi_{ij}(\eta,\chi,\theta,\phi)=\sum_{s=0}^{1} \sum_{n=3}^{+\infty} \sum_{l=2}^{n-1} \sum_{m=-l}^{l}  \pi_{n l m s}(\eta) Q^{n l m s}_{ij}(\chi,\theta,\phi)~.
\ee
The equations for the harmonic components of the tensor perturbations $h_{ij}$ and $H_{ij}$ are then obtained after replacing the decompositions \eqref{Eq:GSmetricperts} and \eqref{Eq:GSmatterpert}
into the evolution equations \eqref{Eq:TensorPert},
\begin{subequations}
\begin{align}
&h^{\prime\prime}_{n l m s} +2 \mathcal{H}\, h^{\prime}_{n l m s}+  \frac{(n^2-3)}{\rnew^2} h_{n l m s} + \frac{2}{\rnew^2}h_{n l m s}+m^2 a^2 \lambda(y) (h_{n l m s}-H_{n l m s}) =\frac{2}{M_g^2} a^2 \pi_{n l m s}~,\\
&H^{\prime\prime}_{n l m s} +\left(2 c\, \mathcal{H}_f -\frac{c^{\prime}}{c}\right) H^{\prime}_{n l m s} + c^2 \frac{(n^2-3)}{\rnew^2} H_{n l m s} + \frac{2 c^2}{\rnew^2}H_{n l m s}+\frac{m^2}{\alpha^2} b^2 c\, y^{-4}\lambda(y) (H_{n l m s}-h_{n l m s}) =0~.
\end{align}
\end{subequations}
In order to make the notation lighter, from this point on, we will drop the labels $n$, $l$, $m$, $s$.
Thus, the dynamics of tensor perturbations is governed by the following two
second-order non-autonomous linear ordinary differential equations,
\begin{subequations}\label{Eq:SystemTensorPerthH}
\begin{align}
&h^{\prime\prime}+2 \mathcal{H}\, h^{\prime}+  \frac{(n^2-1)}{\rnew^2} h+m^2 a^2 \lambda(y) (h-H) =\frac{2}{M_g^2} a^2 \pi~,\label{Eq:TensorPert_Harm1} \\
&H^{\prime\prime} +\left(2 c\, \mathcal{H}_f -\frac{c^{\prime}}{c}\right) H^{\prime} + c^2 \frac{(n^2-1)}{\rnew^2} H +\frac{m^2}{\alpha^2} b^2 c\, y^{-4}\lambda(y) (H-h) =0~.\label{Eq:TensorPert_Harm2}
\end{align}
\end{subequations}

\subsubsection*{Mukhanov-Sasaki-like variables}

In analogy with the standard Mukhanov-Sasaki variables, which are commonly used in general relativity,
it turns out to be very convenient to rescale $h$ and $H$ by their respective scale factors, thus defining the new dynamical variables $\mu\equiv a h$ and $\nu\equiv b H$.
In this way, the system \eqref{Eq:SystemTensorPerthH} can be recast as
\begin{subequations}\label{Eq:TensorPert_Harm_munu}
\begin{align}
&\mu^{\prime\prime}+  \left(k^2 -\frac{a^{\prime\prime}}{a}\right)\mu+m^2 a^2 \lambda(y) (\mu - y^{-1} \nu) =\frac{2}{M_g^2} a^3 \pi~,\label{Eq:TensorPert_Harm12} \\
& \nu^{\prime\prime}+  \left(k^2 -\frac{b^{\prime\prime}}{b}\right)\nu -\frac{c^{\prime}}{c}\left(\frac{\nu^{\prime}}{b}-\frac{b^{\prime}\nu}{b^2}\right)+\frac{m^2}{\alpha^2} b^2 c\,y^{-4}\lambda(y) (\nu - y\, \mu) =0~,\label{Eq:TensorPert_Harm22}
\end{align}
\end{subequations}
where we have defined $k^2\equiv(n^2-1)/\rnew^2$. 

In general, the effects due to a spatially closed universe are most relevant for the lowest lying modes in the spectrum since, as $n$ increases, the difference between consecutive eigenvalues tends to zero.
However, in the next sections we will focus on the analysis of the dynamics of sub-horizon modes (with large
$n$), for which the effects of spatial curvature are unappreciable.

\section{Sub-horizon dynamics of tensor modes in the small $y$-limit}\label{Sec: DynamicsSubHorizon}

In this section we further specify our assumptions on the background evolution, which will be needed to obtain approximate analytical solutions of Eq.~\eqref{Eq:TensorPert_Harm_munu} describing the propagation of gravitational waves in the next sections.
In particular, our aim is to study the dynamics of sub-horizon tensor modes in the GR limit of the theory, specifically in cosmic epochs where the background energy density is dominated by hydrodynamic matter and the ratio $y$ is small.
Our motivation to consider the GR limit is that one expects bimetric effects 
to be small, since GR agrees with observations on all scales. However, even in the regime where departures from GR are strongly suppressed at the background level, the bimetric field equations still give rise to interesting phenomenology for the propagation of tensor perturbations, which is the subject of the following sections.

More precisely,
we will consider an expanding universe (with an initial singularity such that the scale factor $a$ vanishes at $\eta=0$), and assume a barotropic equation of state $p_m=w\,\rho_m$,
with a constant parameter $w>-1$,
for matter. Then, the continuity equation \eqref{Eq:ContinuityEquationMatter} implies that
the matter density behaves as $\rho_m\approx a^{-3(w+1)}$, and thus it diverges at early times $a\rightarrow 0$. Concerning the ratio between scale factors $y$, the quartic equation \eqref{Eq:yAlgebraic} provides four different solutions, which define different branches with
distinctive behaviors at early times. Since three of those branches have been shown to be
unphysical \cite{Luben:2020xll}, we will restrict ourselves to the so-called {\it finite branch} \cite{Konnig:2015lfa}, where $y$ is a monotonically increasing function of $\eta$ and tends to zero as $a\rightarrow 0$ \cite{Hogas:2021lns}. Note that, strictly speaking, for solutions on the finite branch the limit $y\to 0$ is only approached in the very early universe, where the redshift $z$ diverges. Nonetheless, it is possible to satisfy the condition $y\ll 1$ until relatively small redshift values, depending on the initial conditions. For instance, the analysis in Ref.~\cite{Hogas:2021lns} shows that there are viable solutions where $y\ll 1$ as late as $z\sim 100$.

\

Let us thus solve the background equations under the commented assumptions.
For small $y$, Eq.~\eqref{Eq:Densities_Relation} can be approximated as
\begin{equation}\label{Eq:DominantFluid1}
\rho_m=\frac{m^2M_g^2}{\alpha^2}\beta_1 y^{-1}~,
\end{equation}
with $\rho_m>0$ provided that $\beta_1> 0$ \footnote{With our assumptions, from Eq.~\eqref{Eq:BackgroundEqs_f-1} we find that, for $y>0$, $\mathcal{H}^2\to \infty$ as $y\to 0$ only if $\beta_1>0$~\cite{Luben:2020xll}.}.
Differentiating, this gives
\be\label{Eq:DominantFluid2}
\rho_m^{\prime}=-\rho_m\frac{y^{\prime}}{y}~.
\ee
Replacing relation \eqref{Eq:DominantFluid2} in the matter continuity equation \eqref{Eq:ContinuityEquationMatter}, and making use of Eq.~\eqref{Eq:BianchiConstraintRewrite1},
we obtain $c$
in terms of the equation-of-state parameter,
\be\label{Eq:LapseApproxSolution_fin}
c = 1+ 3\left(1+w\right) ~.
\ee

Using the fact that $\rho_m\sim a^{-3(w+1)}$~,
for $y\ll1$ the effective energy density $\rho_g$ can be neglected in Eq. \eqref{Eq:BackgroundEqs_g-1},
whose solution provides the evolution of the background scale factor
\begin{equation}\label{background_solution}
a=a_*\left(\frac{\eta}{\eta_*}\right)^{\frac{2}{3w+1}},
\end{equation}
and the Hubble rate,
\begin{equation}
\mathcal{H}=\frac{2}{3w+1}\eta^{-1}~.
\end{equation}
Substituting this in \eqref{Eq:BianchiConstraintRewrite1}, one obtains
\begin{equation}
\quad \quad y= y_*\left(\frac{\eta}{\eta_*}\right)^{\frac{6(1+w)}{3w+1}},\quad \quad b= b_*\left(\frac{\eta}{\eta_*}\right)^{\frac{8+6w}{3w+1}}~.
\end{equation}
In these expressions $\eta\in[0,\infty)$ and $\eta_*$ is an arbitrary reference value of the conformal time.
Note that all the integration constants have been fixed at the point $\eta_*$, and that the particular case $w=-1/3$ is not included in this solution. The results we will present will thus be completely generic for any constant $w>-1$, excluding $w=-1/3$.

Finally, concerning the tensor perturbations, as already commented, we are going to focus our analysis on sub-horizon modes,
defined by the condition $k\,\eta \gg1$. For simplicity, in the following we will also assume matter with zero anisotropic stress, $\pi=0$.
Using \eqref{background_solution}, it is easy to see that for these modes $a^{\prime\prime}/a\approx 1/\eta^2\ll k^2$~,
and, since $c$ is constant given by \eqref{Eq:LapseApproxSolution_fin},
the system \eqref{Eq:TensorPert_Harm_munu} can be approximated as
\begin{subequations}\label{Eq:TensorPert_munu_subhor}
\begin{align}
&\mu^{\prime\prime}+  k^2 \mu+m^2 a^2 \lambda(y) (\mu - y^{-1} \nu) = 0 ~, \\
&\nu^{\prime\prime}+ c^2 k^2  \nu +\frac{m^2}{\alpha^2} b^2 c\, y^{-4}\lambda(y) (\nu - y\, \mu) = 0 ~.
\end{align}
\end{subequations} 
Our last assumption for the physical scenario under consideration is that the bimetric effects are small,
and thus we are near the GR limit, so that $\alpha\ll 1$. However, this system of equations is still very
involved and it is not possible to obtain exact analytical solutions.
Therefore, in the next section we will consider well-suited methods
to construct approximate analytical solutions.

\

\section{Approximate analytical solutions: multiple-scale analysis}\label{Sec: AnalyticalSolutions}

In order to apply approximate methods, let us first rewrite the system \eqref{Eq:TensorPert_munu_subhor} in a dimensionless form.
In particular, we define a dimensionless time variable $t\equiv k\,\eta$.
Furthermore, we introduce a dimensionless parameter $\epsilon\equiv1/(k\eta_*)$,
where $\eta_*$ will be interpreted as the value of the conformal time $\eta$ at the end of the considered epoch,
where the
small-$y$ assumption still holds.
Since we are assuming sub-horizon modes, $k\eta_*\gg 1$ and thus $\epsilon\ll 1$. However, for the mode to be sub-horizon all along the considered evolution, one would need to define a minimum value of $\eta=\eta_0$, so that $k\eta_0\gg 1$. In this way, the following analysis will in principle be valid for all $\eta$ in the interval $\eta_0\lesssim \eta\lesssim\eta_*$.
Then, after some straightforward algebra, the system~\eqref{Eq:TensorPert_munu_subhor} takes the form
\begin{subequations}\label{Eq:SystemDimless}
\begin{align}
&\frac{\de^2 \mu}{\de t^2}+   (1+\epsilon^2 p(\tilde{t})) \mu - \epsilon^2 q(\tilde{t}) \nu =0~,\label{Eq:SystemDimless_a}\\
&\frac{\de^2 \nu}{\de t^2}+  \left(c^2 +\frac{\epsilon^2}{\alpha^2} r(\tilde{t})\right) \nu - \frac{\epsilon^2}{\alpha^2} s(\tilde{t}) \mu =0~,\label{Eq:SystemDimless_b}
\end{align}
\end{subequations}
where $\tilde{t}\equiv\epsilon\, t$ is a {\it slow time} variable. The time-dependent couplings in Eq.~\eqref{Eq:SystemDimless} are obtained from Eqs.~\eqref{Eq:LambdaDefinitions} and \eqref{Eq:TensorPert_munu_subhor} after substituting the background evolution. Their expressions as functions of $\tilde{t}$ are
\begin{equation}\label{Eq:pqrsDef}
\begin{split}
p(\tilde{t})&=\frac{4}{(1+3w)^2}\left[B_1 y_*\, \tilde{t}^{\frac{10+6w}{1+3w}}+B_2(1+c)y_*^2\, \tilde{t}^{\frac{4(4+3w)}{1+3w}}+B_3 c\, y_*^3\, \tilde{t}^{\frac{22+18w}{1+3w}} \right]~,\\
q(\tilde{t})&=y_* ^{-1}\,\tilde{t}^{\,-\frac{6(1+w)}{1+3w}}p(\tilde{t})~,\\
r(\tilde{t})&=c\, y_* ^{-2}\,\tilde{t}^{\,-\frac{12(1+w)}{1+3w}}p(\tilde{t})~,\\
s(\tilde{t})&=c\, y_* ^{-1}\tilde{t}^{\,-\frac{6(1+w)}{1+3w}}p(\tilde{t})~,
\end{split}
\end{equation}
where we have introduced the dimensionless parameters $B_n\equiv m^2 a_*^2\beta_n/{\cal H}_*^2$ \cite{Mortsell:2017fog}.  

As discussed in Sec.~\ref{Sec:GRlimit}, the  general-relativity limit of bimetric gravity is recovered as $\alpha\to0$ \cite{Hassan:2014vja, Akrami:2015qga}. Thus, the system \eqref{Eq:SystemDimless} naturally involves two small parameters, $\alpha$ and $\epsilon$: the former is fixed at the outset by the values of $M_f$ and $M_g$, whereas the latter is by definition mode-dependent (i.e., for a given mode, it depends on the corresponding eigenvalue of the spatial Laplacian). On closer inspection of the system \eqref{Eq:SystemDimless}, we realize that the behavior of solutions is quite sensitive to the hierarchy satisfied by $\alpha$ and $\epsilon$. In fact, the limit of Eq.~\eqref{Eq:SystemDimless_b} where both $\alpha$ and $\epsilon$ tend to zero clearly depends on the order in which these limits are taken.
Thus, we can identify two different types of modes:
\begin{itemize}
\item Class I: $\epsilon\ll\alpha\ll1$, corresponding to deep sub-horizon modes with $k \eta_{*} \gg M_g/M_f$.
\item Class II: $\alpha\ll\epsilon\ll 1$, corresponding to sub-horizon modes with $1\ll k \eta_*\ll M_g/M_f$.
\end{itemize}
For modes of Class I the couplings are small and evolve over long-time scales determined by $\tilde{t}$. Therefore, the system can be studied using techniques from multiple-scale analysis presented in Ref.~\cite{kevorkian1996multiple}. However, for modes of Class~II by no means can the couplings \eqref{Eq:SystemDimless_b} be considered small, and therefore the methods of Ref.~\cite{kevorkian1996multiple} are not applicable. Such large couplings are responsible for fast superimposed oscillations that require different methods to be appropriately described. These methods will be developed in Sec.~\ref{Sec:case2}.
Finally, modes with $\epsilon\approx\alpha$, that is, with $k\approx M_g/(M_f\eta_*)$, do not lay in any of the mentioned classes.
For such modes, the explicit perturbative parameters in Eq.~\eqref{Eq:SystemDimless_b} are absent, which makes difficult
to design an appropriate perturbative analysis. Ultimately, note that the value of $\epsilon$ is epoch-dependent and gets smaller as the universe moves to later epochs (which have larger values of $\eta_*$). Therefore, Class II modes eventually become Class I in later epochs (as long as the background dynamics is dominated by hydrodynamic matter with $w>-1$).

Before we proceed further with our analysis, it is convenient to recast the system in a different form. We note that Eq.~\eqref{Eq:SystemDimless} has the same structure as the non-autonomous linear system of coupled oscillators considered in Ref.~\cite{kevorkian1996multiple} (see Chapter 4 therein),
\begin{subequations}
\begin{align}
&\frac{\de^2 \mu}{\de t^2}+   \omega_\mu^2(\tilde{t},\epsilon) \mu = \epsilon^2 q(\tilde{t}) \nu~,\\
&\frac{\de^2 \nu}{\de t^2}+  \omega_\nu^2(\tilde{t},\epsilon) \nu  = \epsilon^2 \frac{s(\tilde{t})}{\alpha^2} \mu ~,
\end{align}
\end{subequations}
with $\omega_\mu^2(\tilde{t},\epsilon)=1+\epsilon^2 p(\tilde{t})$ and $\omega_\nu^2(\tilde{t},\epsilon)=c^2+\epsilon^2 r(\tilde{t})/\alpha^2$. This system can then be written in the so-called {\it standard form}, transforming to the action and angle variables defined as
\begin{subequations}
\begin{align}
&{\cal J}_\mu=\frac{\dot{\mu}^2+\omega_\mu^2\, \mu^2}{2\omega_\mu}~,\qquad\qquad \varphi_\mu=\arctan\left(\frac{\omega_\mu \,\mu}{\dot{\mu}}\right)~,\\
&{\cal J}_\nu=\frac{\dot{\nu}^2+\omega_\nu^2 \,\nu^2}{2\omega_\nu}~,\qquad\qquad \varphi_\nu=\arctan\left(\frac{\omega_\nu\, \nu}{\dot{\nu}}\right)~.
\end{align}
\end{subequations}
In terms of these variables, Eqs.~\eqref{Eq:SystemDimless} read,
\begin{subequations}\label{Eq:System_KCform}
\begin{align}
\dot{{\cal J}}_\mu & =-\epsilon \frac{\dot{\omega}_{\mu}}{\omega_{\mu}} {\cal J}_{\mu}\cos(2\varphi_{\mu})+2\epsilon^2q(\tilde{t})\sqrt{\frac{{\cal J}_\mu {\cal J}_\nu} {\omega_{\mu}\omega_{\nu}}} \cos{\varphi_{\mu}}\sin{\varphi_{\nu}}~,\label{Eq:System_KCform_a}\\
\dot{{\cal J}}_\nu & =-\epsilon \frac{\dot{\omega}_{\nu} }{\omega_{\nu}}{\cal J}_{\nu}\cos(2\varphi_{\nu})+2\epsilon^2\frac{s(\tilde{t})}{\alpha^2} \sqrt{\frac{{\cal J}_\mu {\cal J}_\nu} {\omega_{\mu}\omega_{\nu}}}\cos{\varphi_{\nu}}\sin{\varphi_{\mu}}~,\label{Eq:System_KCform_b}\\
\dot{\varphi}_\mu & =\omega_{\mu}+\epsilon\frac{ \dot{\omega}_{\mu} }{2\omega_{\mu}}\sin(2\varphi_{\mu})-\epsilon^2q(\tilde{t})\sqrt{\frac{{\cal J}_\nu}{{\cal J}_\mu}}\frac{  \sin{\varphi_{\mu}}\sin{\varphi_{\nu}}}{\sqrt{\omega_{\mu}\omega_{\nu}}}~,\label{Eq:System_KCform_c}\\
\dot{\varphi}_\nu & =\omega_{\nu}+\epsilon \frac{\dot{\omega}_{\nu}}{2\omega_{\nu}} \sin(2\varphi_{\nu})-\epsilon^2 \frac{s(\tilde{t})}{\alpha^2}\sqrt{\frac{{\cal J}_\mu}{{\cal J}_\nu}}\frac{ \sin{\varphi_{\mu}}\sin{\varphi_{\nu}}}{\sqrt{\omega_{\mu}\omega_{\nu}}}~.\label{Eq:System_KCform_d}
\end{align}
\end{subequations}
This is indeed the form of the system we will consider to obtain approximate solutions.

\subsection{Class I}
We will begin analyzing modes of Class I, which have $\epsilon\ll\alpha\ll1$. The asymptotics of the solutions of system \eqref{Eq:System_KCform} in this regime is controlled solely by $\epsilon$. We observe that the system \eqref{Eq:System_KCform} belongs to the more general class
\begin{subequations}\label{Eq:System_KCform_general}
\begin{align}
\dot{{\cal J}}_N & = \epsilon\, F_N({\cal J}_i,\varphi_i,\tilde{t};\epsilon) ~,\\
\dot{\varphi}_N & = \omega^{(0)}_N({\cal J}_i,\tilde{t})+\epsilon\, G_N({\cal J}_i,\varphi_i,\tilde{t};\epsilon)~.
\end{align}
\end{subequations}
In our particular case, we have $N\in \{\mu,\nu\}$ for the labels, while the unperturbed frequencies are constant and given by $\omega^{(0)}_\mu=1$, $\omega^{(0)}_\nu=c$, and the remaining functions $F_N$, $G_N$ read
\begin{subequations}\label{Eq:Def_FG}
\begin{align}
F_\mu({\cal J}_i,\varphi_i,\tilde{t};\epsilon) &= -\frac{\dot{\omega}_{\mu}}{\omega_{\mu}} {\cal J}_{\mu}\cos(2\varphi_{\mu})+2\epsilon\, q(\tilde{t})\sqrt{\frac{{\cal J}_\mu {\cal J}_\nu} {\omega_{\mu}\omega_{\nu}}} \cos{\varphi_{\mu}}\sin{\varphi_{\nu}}~,\\
F_\nu({\cal J}_i,\varphi_i,\tilde{t};\epsilon) &= -\frac{\dot{\omega}_{\mu} }{\omega_{\nu}}{\cal J}_{\nu}\cos(2\varphi_{\nu})+\epsilon\frac{2 s(\tilde{t})}{\alpha^2} \sqrt{\frac{{\cal J}_\mu {\cal J}_\nu} {\omega_{\mu}\omega_{\nu}}}\cos{\varphi_{\nu}}\sin{\varphi_{\mu}}~,\\
G_\mu({\cal J}_i,\varphi_i,\tilde{t};\epsilon)&=\frac{\omega_{\mu}-\omega^{(0)}_\mu}{\epsilon}+ \frac{ \dot{\omega}_{\mu} }{2\omega_{\mu}}\sin(2\varphi_{\mu})-\epsilon\, q(\tilde{t})\sqrt{\frac{{\cal J}_\nu}{{\cal J}_\mu}}\frac{  \sin{\varphi_{\mu}}\sin{\varphi_{\nu}}}{\sqrt{\omega_{\mu}\omega_{\nu}}}~,\\
G_\nu({\cal J}_i,\varphi_i,\tilde{t};\epsilon)&=\frac{\omega_{\nu}-\omega^{(0)}_\nu}{\epsilon}+ \frac{\dot{\omega}_{\nu}}{2\omega_{\nu}} \sin(2\varphi_{\nu})-\epsilon \frac{s(\tilde{t})}{\alpha^2}\sqrt{\frac{{\cal J}_\mu}{{\cal J}_\nu}}\frac{ \sin{\varphi_{\mu}}\sin{\varphi_{\nu}}}{\sqrt{\omega_{\mu}\omega_{\nu}}}~,
\end{align}
\end{subequations}
where we have stressed the dependence on $\tilde{t}$. Moreover, we observe that the condition $\epsilon\ll\alpha\ll1$ implies that \eqref{Eq:System_KCform} describes, in the language of Ref.~\cite{kevorkian1996multiple}, a standard-form system. In order to apply the techniques developed in Ref.~\cite{kevorkian1996multiple}, based on the method of multiple scales \cite{bender1999multiple}, to obtain asymptotic solutions for this kind of systems, the function $F_N$ and $G_N$ in \eqref{Eq:System_KCform_general} must satisfy the following two conditions: (i) all the functions $F_N$ and $G_N$ are $O(1)$ in the $\epsilon\to0$ limit; (ii) $F_N$ and $G_N$ are periodic functions of the $\varphi_i$ with period $2\pi$. Note that, for condition (i) to follow by our functions \eqref{Eq:Def_FG}, we must require the leading order in $r(\tilde{t})$, proportional to $\tilde{t}^{-2}$, to remain sufficiently small during all over the time interval we are considering. This imposes $B_1\sim\tilde{t}_0{}^2\alpha^2y_*(1+3w)^2/4c$.

The method of multiple scales is able to account for the cumulative effect of small perturbations over long timescales, of order $t\sim\epsilon^{-1}$. Thus, the effect of perturbations can be understood as a further dependence of the dynamics on a {\it slow time} variable $\tilde{t}$, in addition to the {\it fast time} $t$ of the unperturbed system. The method of multiple scales treats $t$ and $\tilde{t}$ as independent, exploiting the additional freedom to remove secular terms (which would grow monotonically in time) and thus achieve a uniform perturbative expansion over long timescales.

The dependence of the dynamics on the timescales $t$ and $\tilde{t}$, along with the smallness of $\epsilon$, motivates the following perturbative ansatz for the solution of \eqref{Eq:System_KCform}, 
\begin{subequations}\label{Eq:Ansatz_case1}
\begin{align}
{\cal J}_N(t;\epsilon)&={\cal J}_N^{(0)}(\tilde{t})+\epsilon\, {\cal J}_N^{(1)}(\tau_i,\tilde{t})+\epsilon^2{\cal J}_N^{(2)}(\tau_i,\tilde{t})+\epsilon^3{\cal J}_N^{(3)}(\tau_i,\tilde{t})+\mathcal{O}(\epsilon^4)~,\\
\varphi_N(t;\epsilon)&=\varphi_N^{(0)}(\tau_N,\tilde{t})+\epsilon\, \varphi_N^{(1)}(\tau_i,\tilde{t})+\epsilon^2 \varphi_N^{(2)}(\tau_i,\tilde{t})+\epsilon^3 \varphi_N^{(3)}(\tau_i,\tilde{t})+\mathcal{O}(\epsilon^4)~.
\end{align}
\end{subequations}
The $\tau_i$ are the fast times associated with $\munderbar{\omega}_N$ frequencies. As such, they are in principle slowly varying, and thus can be formally defined as
\be
\frac{\de \tau_N}{\de t}=\Omega_N(\tilde{t})~,
\ee
where the quantities $\Omega_N$ are functions of $\tilde{t}$ to be determined later on. Thus, treating the time variables $\tau_i$ and $\tilde{t}$ as independent, we can represent the total time derivative as
\be
\frac{\de }{\de t}=\Omega_\mu(\tilde{t})\frac{\pa }{\pa \tau_\mu}+\Omega_\nu(\tilde{t})\frac{\pa }{\pa \tau_\nu}+\epsilon \frac{\partial }{\partial \tilde{t}}~.
\ee
This is the central equation underlying the method of multiple scales. Note that to zero-th order, $\varphi_N$ depends on $\tau_N$~---although not on the remaining $\tau_i$ with $i\neq N$
\footnote{In fact, it can be shown that a more general ansatz where $\varphi^{(0)}_N$ {\it a priori} depends on all $\tau_i$ gives unwanted secular terms. The condition for the elimination of such secular terms then leads to $\varphi_N^{(0)}(\tau_i,\tilde{t})=\varphi_N^{(0)}(\tau_N,\tilde{t})$ \cite{kevorkian1996multiple}.}.

We substitute the ansatz \eqref{Eq:Ansatz_case1} into Eq.~\eqref{Eq:System_KCform_general}, expand in powers of $\epsilon$ and collect the contributions to equal order in $	\epsilon$. The next step is 
to decompose the functions $F_N$ and $G_N$ into their average and oscillatory parts (the latter are defined as having zero average with respect to all the $\varphi_i$). 
When solving the averaged equations at each order, we must ensure that our solutions are free of secular contributions, following the steps outlined below. Explicitly, the decomposition of $F_N(\mathcal{J}_i,\varphi_i,\tilde{t};\epsilon)$, that is periodic in each  $\varphi_\mu$ and $\varphi_\nu$, is decomposed into an {\it average} term $\overline{F}_N(\mathcal{J}_i,\tilde{t};\epsilon)$ and an {\it oscillatory} term $\widehat{F}_N(\mathcal{J}_i,\varphi_i,\tilde{t};\epsilon)$:
\begin{equation}
F_N(\mathcal{J}_i,\varphi_i,\tilde{t};\epsilon)=\overline{F}_N(\mathcal{J}_i,\tilde{t};\epsilon)+\widehat{F}_N(\mathcal{J}_i,\varphi_i,\tilde{t};\epsilon)~,
\end{equation}
where
\begin{equation}
\overline{F}_N(\mathcal{J}_i,\tilde{t};\epsilon)\equiv\frac{1}{(2\pi)^2}\int_0^{2\pi}\int_0^{2\pi}F_N(\mathcal{J}_i,\varphi_i,\tilde{t};\epsilon)\,\de \varphi_\mu\de \varphi_\nu~,
\end{equation}
and, thus, the oscillatory part can be computed as $\widehat{F}_N(\mathcal{J}_i,\varphi_i,\tilde{t};\epsilon)= F_N(\mathcal{J}_i,\varphi_i,\tilde{t};\epsilon)-\overline{F}_N(\mathcal{J}_i,\tilde{t};\epsilon)$~.
The decomposition of $G_N$ is entirely analogous. Next, each function also needs to be expanded in powers of $\epsilon$. Applying the steps we just outlined to the particular case represented by Eq.~\eqref{Eq:Def_FG}, we obtain
\begin{subequations}
\begin{align}
\overline{F}_N(\mathcal{J}_i,\tilde{t};\epsilon) & =0~,\\
\overline{G}_N(\mathcal{J}_i,\tilde{t};\epsilon) & =\frac{\omega_N-\omega^{(0)}_N}{\epsilon}=\epsilon\, \overline{G}_N^{(1)}(\tilde{t})+\mathcal{O}(\epsilon^{3})~,\\
\widehat{F}_N(\mathcal{J}_i,\varphi_i,\tilde{t};\epsilon) & =\epsilon\,\widehat{F}_{N}^{ (1)}(\mathcal{J}_i^{(0)},\varphi_i^{(0)} ,\tilde{t})+\epsilon^2\widehat{F}_{N}^{(2)}(\mathcal{J}_i^{(0)} ,\varphi_i^{(0)} ,\mathcal{J}_i^{(1)} ,\varphi_i^{(1)} ,\tilde{t})+\mathcal{O}(\epsilon^{3})~,\\
\widehat{G}_{N}(\mathcal{J}_i,\varphi_i,\tilde{t};\epsilon) & =\epsilon\,\widehat{G}_{N}^{(1)}\!(\mathcal{J}_i^{(0)} ,\varphi_i^{(0)},\tilde{t})+\epsilon^2\widehat{G}_{N}^{ (2)}(\mathcal{J}_i^{(0)},\varphi_i^{(0)},\mathcal{J}_i^{(1)},\varphi_i^{(1)},\tilde{t})+\mathcal{O}(\epsilon^{3})~.
\end{align}
\end{subequations}
We observe that at order $\epsilon^2$ all the contributions come from the oscillatory terms. This way, we obtain the following differential equations for equal powers of $\epsilon$:
\begin{flalign}\label{Eq:zeroth}
\mathcal{O}(1):\quad\frac{\partial \varphi_N^{(0)}}{\partial \tau_N}\Omega_N=\omega^{(0)}_N~,\hspace{4.8cm}
\end{flalign}
\begin{subequations}\label{Eq:first}
\begin{flalign}
\mathcal{O}(\epsilon):\quad & \frac{\de \mathcal{J}_N^{(0)}}{\de \tilde{t}}+\Omega_\mu\frac{\partial \mathcal{J}_N^{(1)}}{\partial \tau_\mu}+\Omega_\nu\frac{\partial \mathcal{J}_N^{(1)}}{\partial \tau_\nu}=0~,\hspace{1.55cm}\label{Eq:first_a}\\
& \frac{\partial \varphi_N^{(0)}}{\partial \tilde{t}}+\Omega_\mu\frac{\partial \varphi_N^{(1)}}{\partial \tau_\mu}+\Omega_\nu\frac{\partial \varphi_N^{(1)}}{\partial \tau_\nu}=0~,\label{Eq:first_b}
\end{flalign}
\end{subequations}
\begin{subequations}\label{Eq:second}
\begin{flalign}
\mathcal{O}(\epsilon^2):\quad & \frac{\partial \mathcal{J}_N^{(1)}}{\partial \tilde{t}}+\Omega_\mu\frac{\partial \mathcal{J}_N^{(2)}}{\partial \tau_\mu}+\Omega_\nu\frac{\partial \mathcal{J}_N^{(2)}}{\partial \tau_\nu}=\widehat{F}_N^{(1)}~,\\
& \frac{\partial \varphi_N^{(1)}}{\partial \tilde{t}}+\Omega_\mu\frac{\partial \varphi_N^{(2)}}{\partial \tau_\mu}+\Omega_\nu\frac{\partial \varphi_N^{(2)}}{\partial \tau_\nu}=\overline{G}_N^{(1)}+\widehat{G}_N^{(1)}~,
\end{flalign}
\end{subequations}
\begin{subequations}\label{Eq:third}
\begin{flalign}
\mathcal{O}(\epsilon^3):\quad & \frac{\partial \mathcal{J}_N^{(2)}}{\partial \tilde{t}}+\Omega_\mu\frac{\partial \mathcal{J}_N^{(3)}}{\partial \tau_\mu}+\Omega_\nu\frac{\partial \mathcal{J}_N^{(3)}}{\partial \tau_\nu}=\widehat{F}_N^{(2)}~,\\
& \frac{\partial \varphi_N^{(2)}}{\partial \tilde{t}}+\Omega_\mu\frac{\partial \varphi_N^{(3)}}{\partial \tau_\mu}+\Omega_\nu\frac{\partial \varphi_N^{(3)}}{\partial \tau_\nu}=\widehat{G}_N^{(2)}~.\hspace{1.25cm}
\end{flalign}
\end{subequations}

In order to solve the previous equations avoiding secular terms in the solutions, at each step we must remove those contributions leading to terms that grow with the fast times $\tau_\mu$ and $\tau_\nu$ and would otherwise be responsible for the breakdown of perturbation theory over time scales of order $1/\epsilon$. We illustrate the procedure concretely for the zero-th order solutions. From \eqref{Eq:zeroth} we get
\begin{equation}
\varphi_N^{(0)}(\tau_N,\tilde{t})=d_N(\tilde{t})\,\tau_N+f_{\varphi_N}^{(0)}(\tilde{t})~,\quad \text{with}\quad d_N(\tilde{t})=\frac{\omega^{(0)}_N}{\Omega_N(\tilde{t})}.
\end{equation}
To determine $f_{\varphi_N}^{(0)}$ we plug the obtained solution for $\varphi_N^{(0)}$ in \eqref{Eq:first_b},
\begin{equation}\label{Eq:first_b_secular}
{d_N}^{\prime}(\tilde{t})\,\tau_N+f_{\varphi_N}^{(0)}{}^{\prime}(\tilde{t})+\Omega_\mu\frac{\partial \varphi_N^{(1)}}{\partial \tau_\mu}+\Omega_\nu\frac{\partial \varphi_N^{(1)}}{\partial \tau_\nu}=0~.
\end{equation}
We observe that the solution for $\varphi_N^{(1)}$ would involve quadratic terms in $\tau_N$ unless we impose ${d_N}^{\prime}(\tilde{t})=0$. We can then set $d_N=1$ with no loss of generality. Hence, 
\be
\Omega_{\mu}=\omega^{(0)}_\mu=1 ~, \quad \Omega_{\nu}=\omega^{(0)}_\nu=c~,
\ee
and therefore we have $\tau_\mu=t$ and $\tau_\nu=c\,t$. On the other hand, we remove the averaged term in \eqref{Eq:first_b_secular} by setting $f_{\varphi_N}^{(0)}{}^{\prime}(\tilde{t})=0$, which gives $f_{\varphi_N}^{(0)}(\tilde{t})=\sigma_N^{(0)}$. The solution for $\mathcal{J}_N^{(0)}$ is directly obtained from \eqref{Eq:first_a} by requiring the vanishing of secular terms in  $\mathcal{J}_N^{(1)}$, that is, $\mathcal{J}_N^{(0)}{}^{\prime}(\tilde{t})=0$, and then $\mathcal{J}_N^{(0)}(\tilde{t})=\gamma_N^{(0)}$. Thus, the zero-th order solution reads
\begin{subequations}\label{Eq: ZerothOrderSolcaseI}
\begin{align}
&{\cal J}_\mu^{(0)}(t;\epsilon)=\gamma_\mu^{(0)}~,\hspace{1cm} {\cal J}_\nu^{(0)}(t;\epsilon)=\gamma_\nu^{(0)}~,\\
&\varphi_\mu^{(0)}(t;\epsilon)=t+\sigma_\mu^{(0)}~,\quad \;\varphi_\nu^{(0)}(t;\epsilon)=c\,t+\sigma_\nu^{(0)}~.
\end{align}
\end{subequations}
Note that for the system at hand also the phases $\sigma_i^{(0)}$ are independent of $\tilde{t}$, since the frequencies $\omega^{(0)}_N$ are constant and $\overline{G}_N^{(0)}=0$~; thus, they merely amount to integration constants.

The first-order corrections are obtained by solving the first and second-order equations. Specifically, the first-order equations imply that
\begin{subequations}\label{Eq: FirstOrderSolcaseI}
\begin{align}
&{\cal J}_\mu^{(1)}(t;\epsilon)=f_{\mathcal{J}_{\mu}}^{(1)}(\tilde{t})~,\quad {\cal J}_\nu^{(1)}(t;\epsilon)=f_{\mathcal{J}_{\nu}}^{(1)}(\tilde{t})~,\\
&\varphi_\mu^{(1)}(t;\epsilon)=f_{\varphi_{\mu}}^{(1)}(\tilde{t})~,\;\,\quad \varphi_\nu^{(1)}(t;\epsilon)=f_{\varphi_{\nu}}^{(1)}(\tilde{t})~.
\end{align}
\end{subequations}
Then, the slowly evolving functions on the right-hand sides of Eq.~\eqref{Eq: FirstOrderSolcaseI} are determined by the second-order equations, which give
\be
f_{\mathcal{J}_{\mu}}^{(1)}(\tilde{t})=\gamma_\mu^{(1)}  ~,\quad f_{\mathcal{J}_{\nu}}^{(1)}(\tilde{t})=\gamma_\nu^{(1)}~,
\ee
and
\be
f_{\varphi_{\mu}}^{(1)}(\tilde{t})=\frac{1}{2}\int \de\tilde{t}\, p(\tilde{t}) ~, \quad f_{\varphi_{\nu}}^{(1)}(\tilde{t})=\frac{1}{2 c \alpha ^2}\int \de\tilde{t}\, r(\tilde{t})~.
\ee
The coupling between the two oscillators only has an effect starting from the second order in the perturbative expansion. Following analogous steps as above, we solve the second- and third-order equations to obtain
\begin{subequations}\label{Eq: SecondOrderSolcaseI}
\begin{align}
&\varphi_\mu^{(2)}(t;\epsilon)=\sigma_\mu^{(2)}-\frac{q(\tilde{t})}{2}\left(\frac{\gamma_\nu^{(0)}}{c\,\gamma_\mu^{(0)}}\right)^{\! \! 1/2}\left[\frac{\sin\left(\sigma_\mu^{(0)}-\sigma_\nu^{(0)}+(1-c)t\right)}{1-c}-\frac{\sin\left(\sigma_\mu^{(0)}+\sigma_\nu^{(0)}+(1+c)t\right)}{1+c}\right]~,\\
&\varphi_\nu^{(2)}(t;\epsilon)=\sigma_\nu^{(2)}-\frac{s(\tilde{t})}{2 \alpha ^2}\left(\frac{\gamma_\mu^{(0)}}{c\,\gamma_\nu^{(0)}}\right)^{\! \! 1/2}\left[\frac{\sin\left(\sigma_\nu^{(0)}-\sigma_\mu^{(0)}-(1-c)t\right)}{c-1}-\frac{\sin\left(\sigma_\nu^{(0)}+\sigma_\mu^{(0)}+(c+1)t\right)}{c+1}\right]~,\\
&{\cal J}_\mu^{(2)}(t;\epsilon)=\gamma_\mu^{(2)}+q(\tilde{t})\left(\frac{\gamma_\mu^{(0)}\gamma_\nu^{(0)}}{c}\right)^{\! \! 1/2}\left[\frac{\cos\left(\sigma_\mu^{(0)}-\sigma_\nu^{(0)}+(1-c)t\right)}{1-c}-\frac{\cos\left(\sigma_\mu^{(0)}+\sigma_\nu^{(0)}+(1+c)t\right)}{1+c}\right]~,\\
&{\cal J}_\nu^{(2)}(t;\epsilon)=\gamma_\nu^{(2)}+\frac{s(\tilde{t})}{\alpha^2}\left(\frac{\gamma_\nu^{(0)}\gamma_\mu^{(0)}}{c}\right)^{\! \! 1/2}\left[\frac{\cos\left(\sigma_\nu^{(0)}-\sigma_\mu^{(0)}-(1-c)t\right)}{c-1}-\frac{\cos\left(\sigma_\nu^{(0)}+\sigma_\mu^{(0)}+(c+1)t\right)}{c+1}\right]~.
\end{align}
\end{subequations}

\begin{figure}[hbtp]
 \begin{subfigure}{0.5\textwidth}
        \centering
       \includegraphics[scale=0.18]{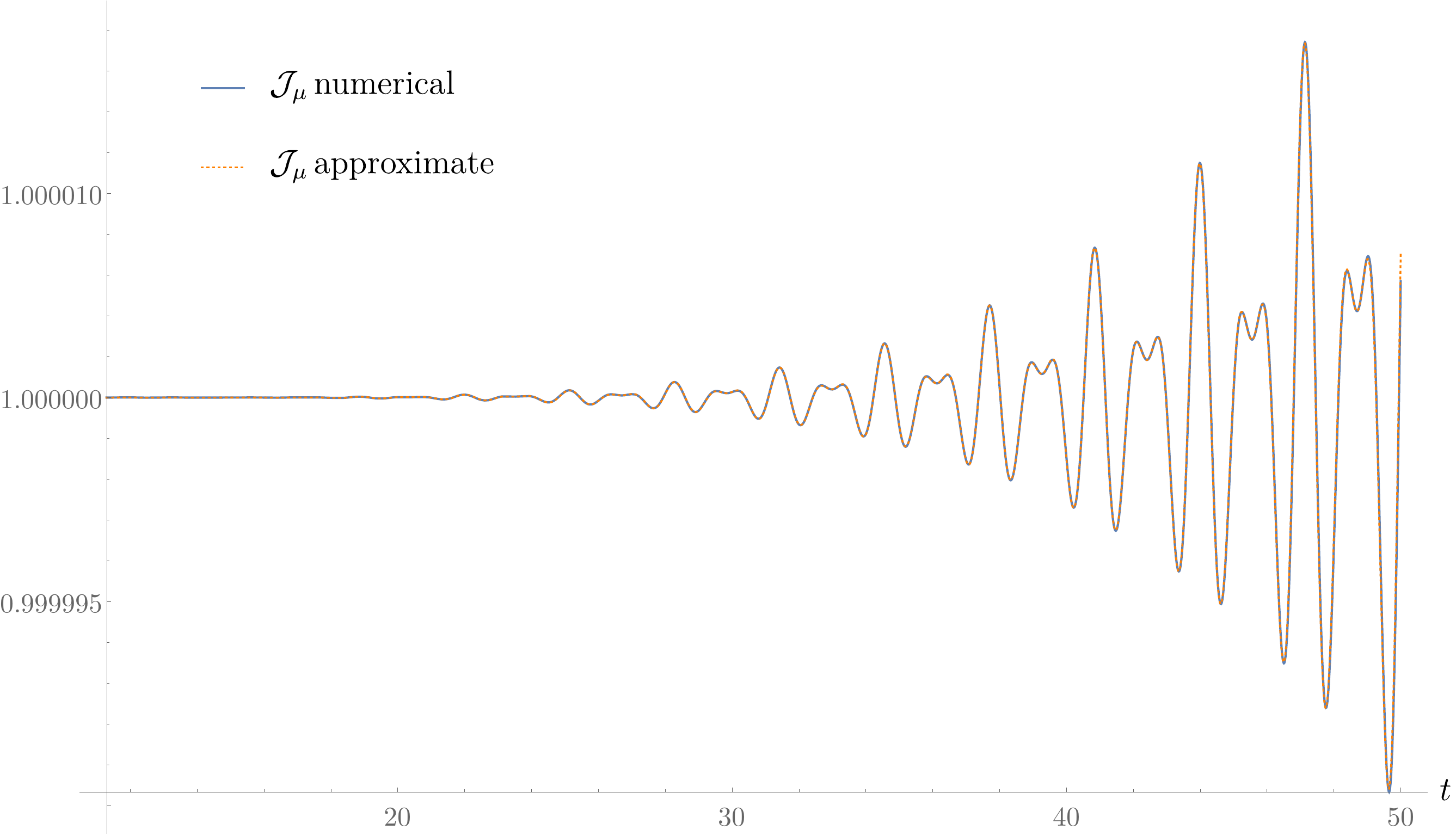}
        \subcaption{}
    \end{subfigure}
     \begin{subfigure}{0.5\textwidth}
        \centering
       \includegraphics[scale=0.173]{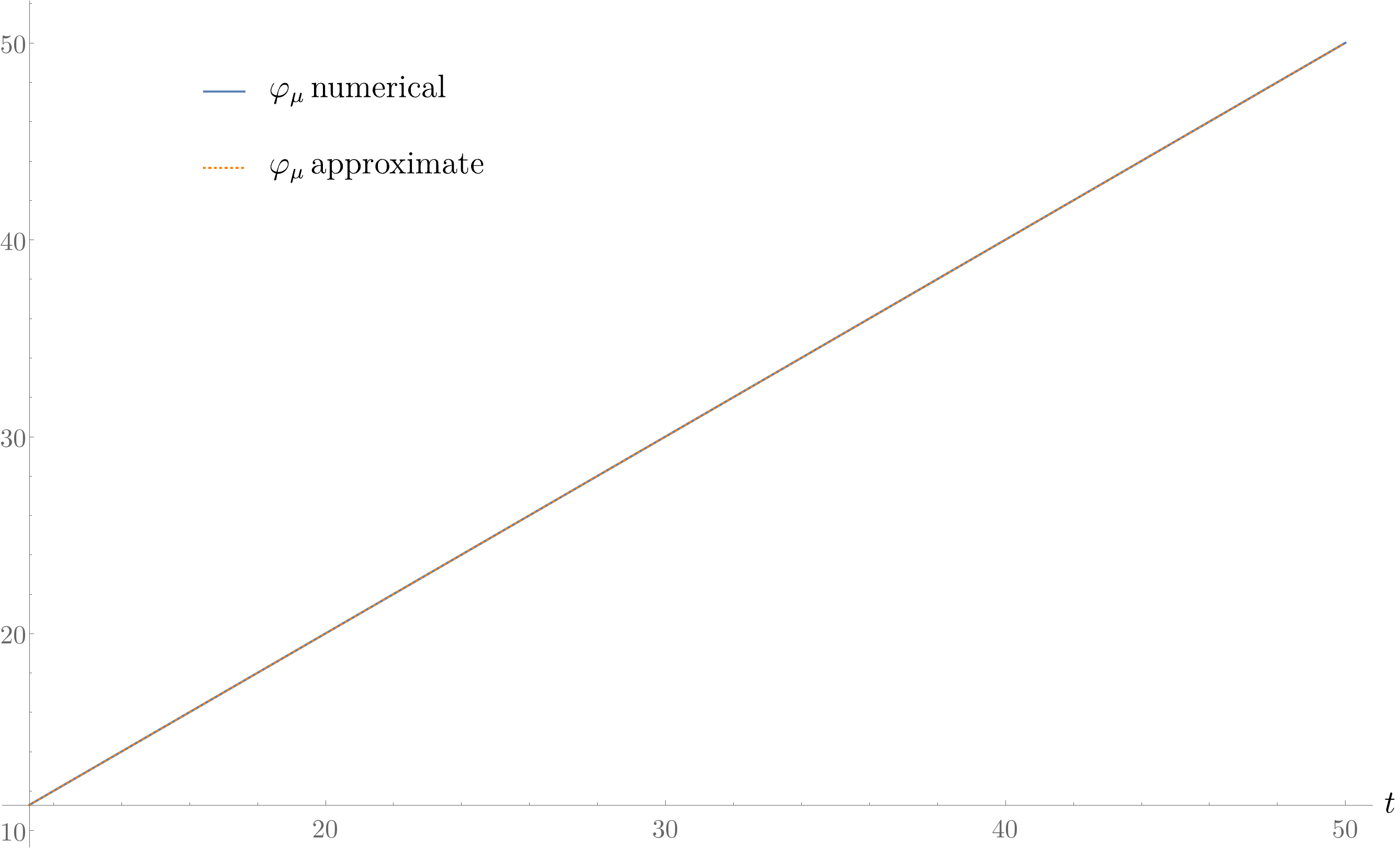}
        \subcaption{}
    \end{subfigure}
     \begin{subfigure}{0.5\textwidth}
        \centering
       \includegraphics[scale=0.18]{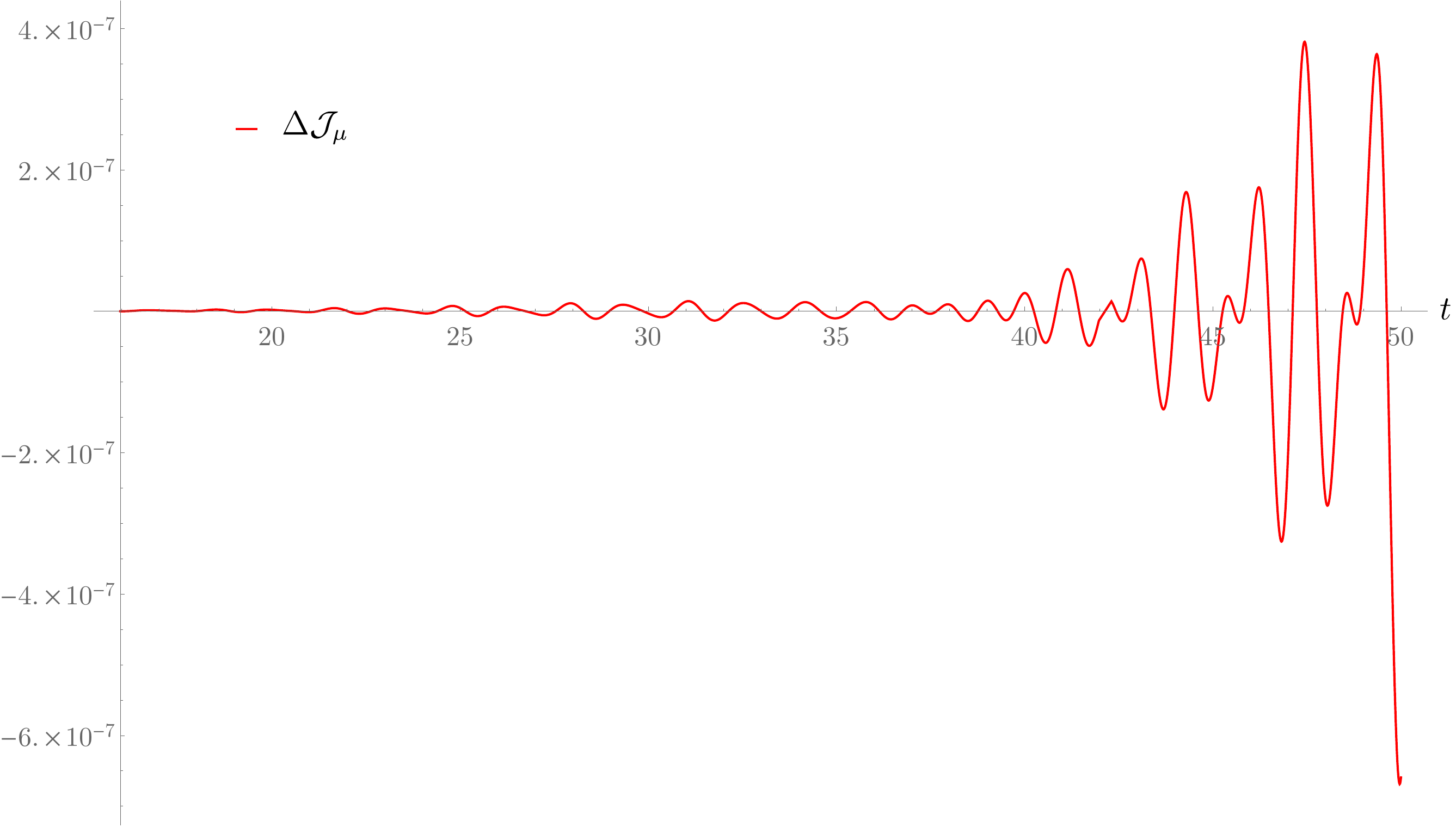}
        \subcaption{}
    \end{subfigure}
    \begin{subfigure}{0.5\textwidth}
        \centering
       \includegraphics[scale=0.18]{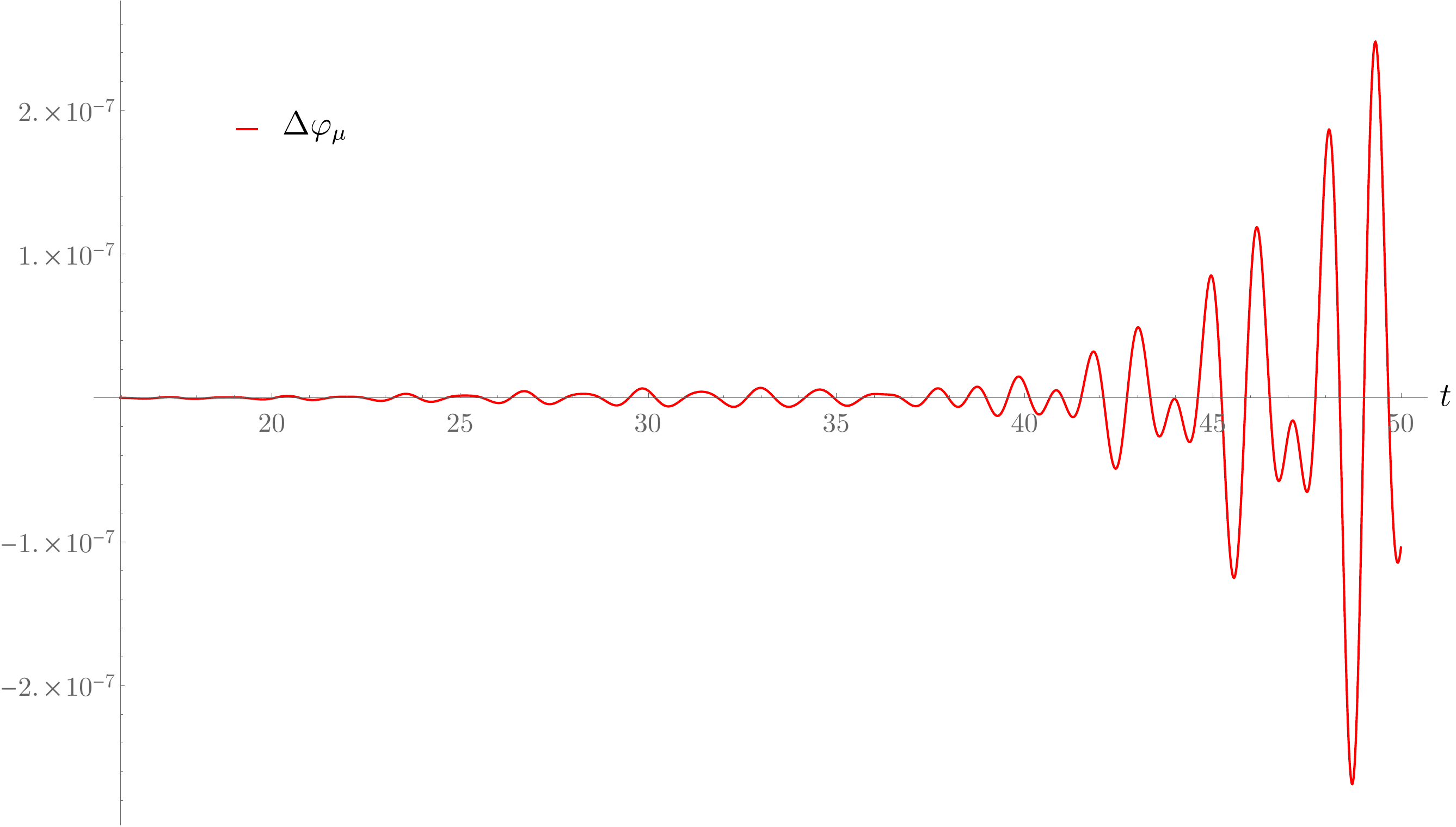}
        \subcaption{}
    \end{subfigure}
\caption{Evolution of the approximate solution obtained in \eqref{Eq: ZerothOrderSolcaseI}, \eqref{Eq: FirstOrderSolcaseI} and \eqref{Eq: SecondOrderSolcaseI}, and the exact numerical solution of \eqref{Eq:System_KCform} for (a) $\mathcal{J}_{\mu}$ and (b) $\varphi_{\mu}$ for Class I modes during the radiation dominated era. (c) and (d) show the difference between the numerical and the approximate solutions, where $\Delta \mathcal{J}_{\mu}\equiv \mathcal{J}_{\mu \text{\,numerical}}-\mathcal{J}_{\mu \text{\,approximate}}$ and $\Delta \varphi_{\mu}\equiv \varphi_{\mu \text{\,numerical}}-\varphi_{\mu \text{\,approximate}}$. For these simulations, we have chosen $\epsilon=0.02,\, \alpha=0.7,\, y_0=0.01,\, B_1= 10^{-3},\, B_2= -2,\, B_3= 3,\, \gamma_\mu^{(0)}= 1,\,\gamma_\nu^{(0)}=2,\,\sigma_ \mu^{(0)}=0,\,\sigma_\nu^{(0)}= 0,\,   \sigma_ \mu^{(1)}=0,\,\sigma_ \mu^{(2)}=0 ,\,\sigma_ \nu^{(1)}=0,\,\sigma_ \nu^{(2)}=0,\,\gamma_\mu^{(1)}= 0,\,\gamma_\mu^{(2)}= 0,\,\gamma_\nu^{(1)}= 0,\,\gamma_\nu^{(2)}= 0$}\label{Fig:caseI-mu}
\end{figure}

\begin{figure}[hbtp]
 \begin{subfigure}{0.5\textwidth}
        \centering
       \includegraphics[scale=0.18]{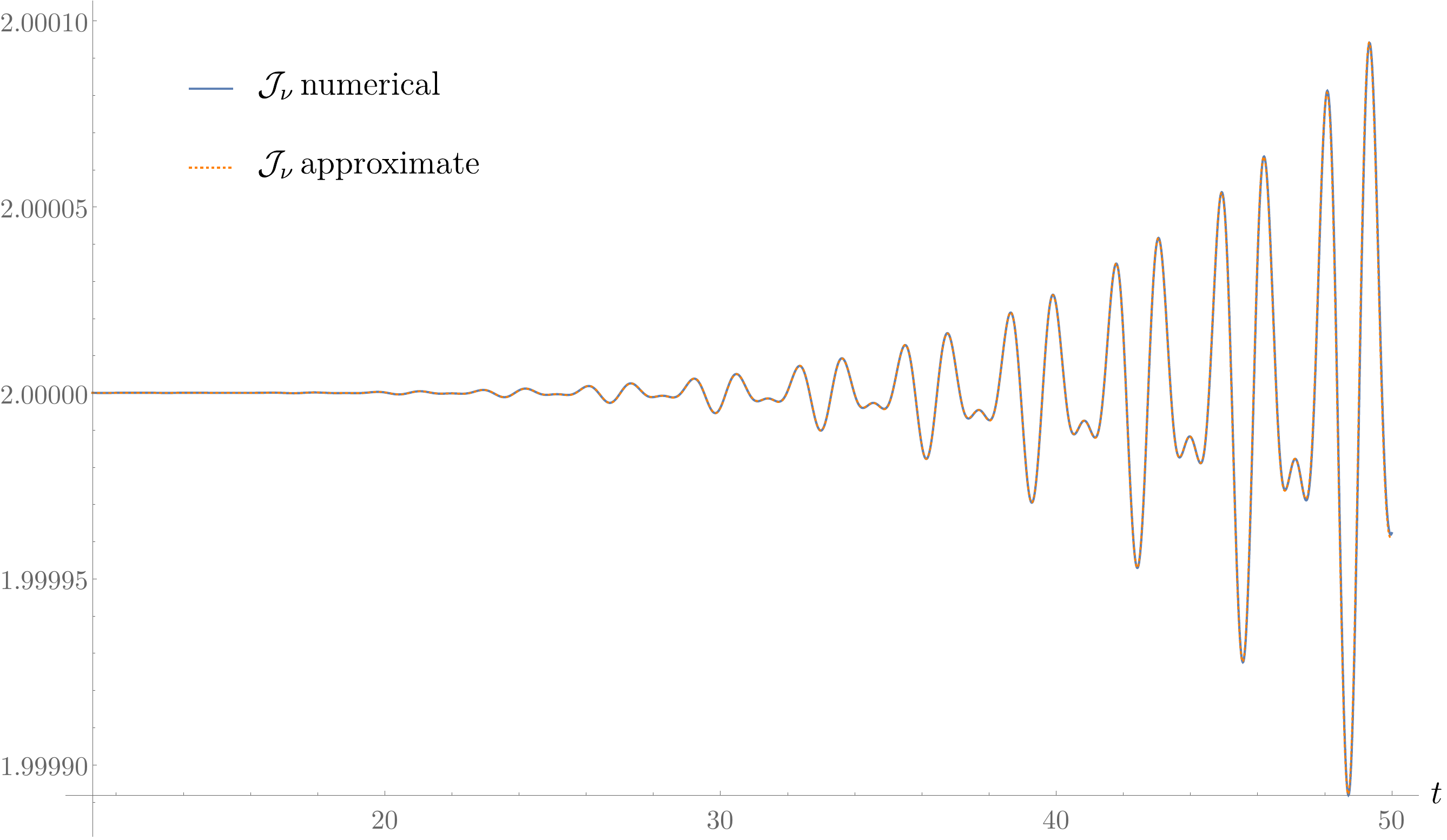}
        \subcaption{}
    \end{subfigure}
     \begin{subfigure}{0.5\textwidth}
        \centering
       \includegraphics[scale=0.173]{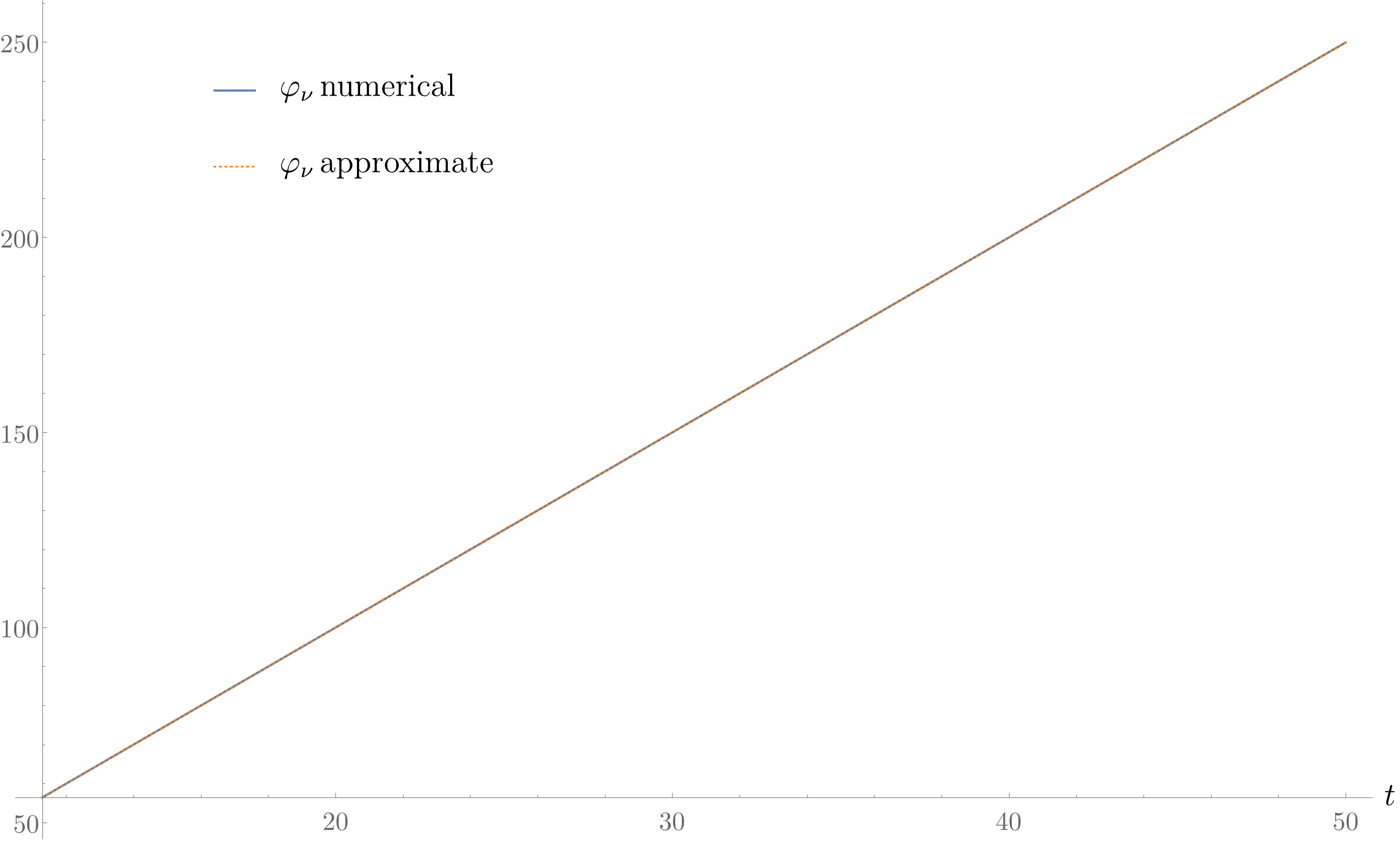}
        \subcaption{}
    \end{subfigure}
     \begin{subfigure}{0.5\textwidth}
        \centering
       \includegraphics[scale=0.18]{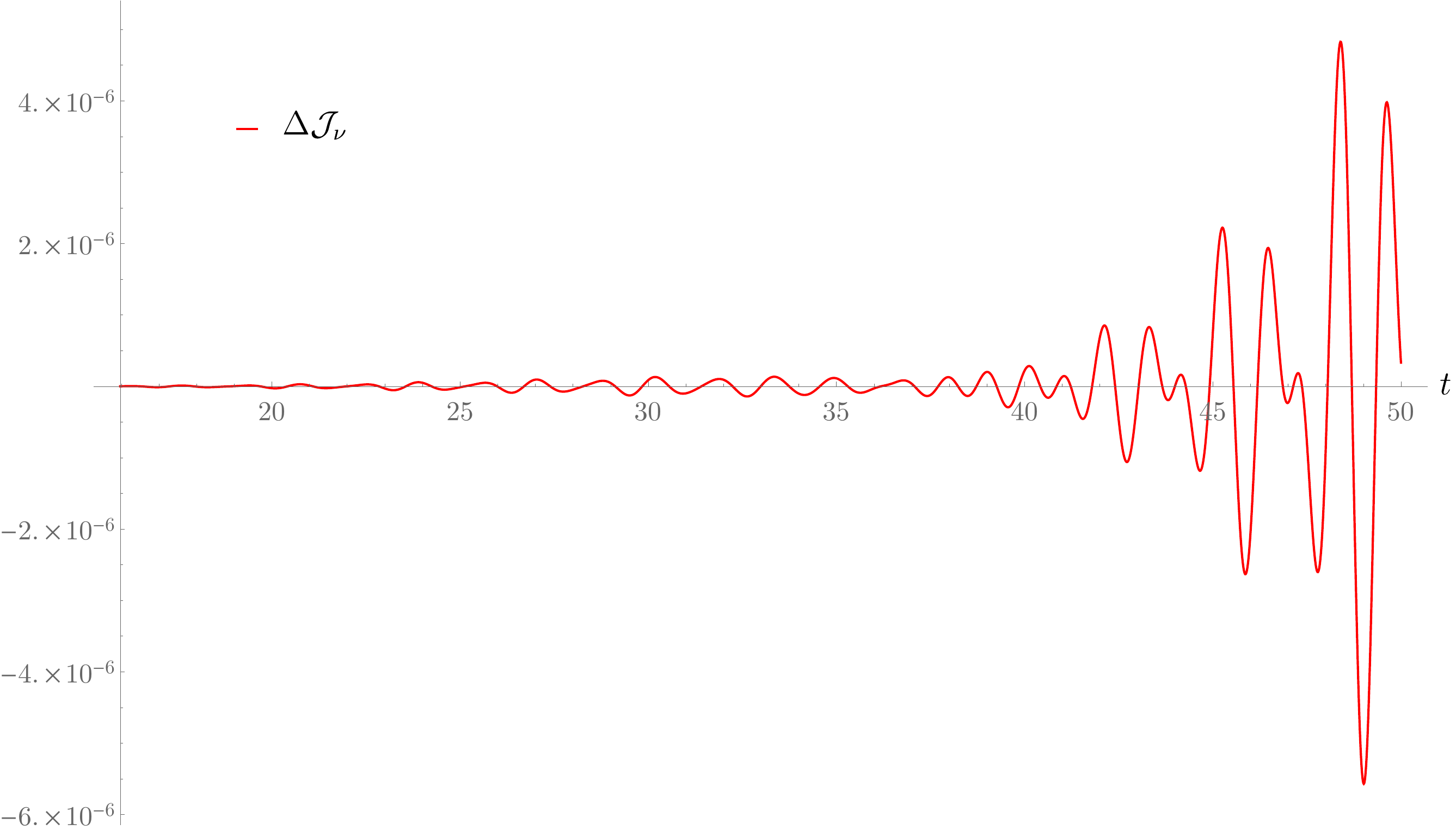}
        \subcaption{}
    \end{subfigure}
    \begin{subfigure}{0.5\textwidth}
        \centering
       \includegraphics[scale=0.18]{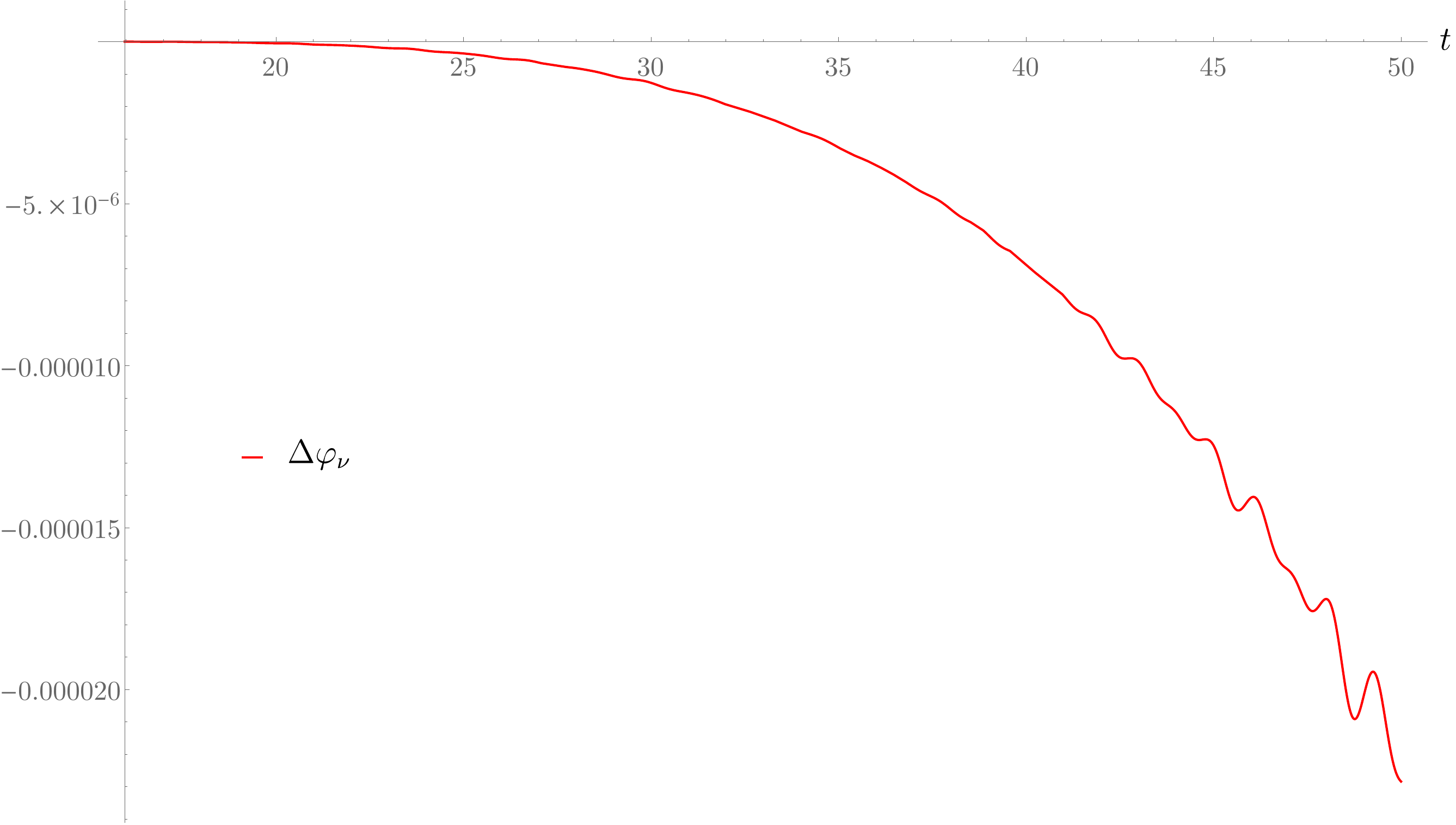}
        \subcaption{}
    \end{subfigure}
\caption{Evolution of the approximate solution obtained in \eqref{Eq: ZerothOrderSolcaseI}, \eqref{Eq: FirstOrderSolcaseI} and \eqref{Eq: SecondOrderSolcaseI}, and the exact numerical solution of \eqref{Eq:System_KCform} for (a) $\mathcal{J}_{\nu}$ and (b) $\varphi_{\nu}$ for Class I modes during the radiation dominated era. (c) and (d) show the difference between the numerical and the approximate solutions, where $\Delta \mathcal{J}_{\nu}\equiv \mathcal{J}_{\nu \text{\,numerical}}-\mathcal{J}_{\nu \text{\,approximate}}$ and $\Delta \varphi_{\nu}\equiv \varphi_{\nu \text{\,numerical}}-\varphi_{\nu \text{\,approximate}}$. For these simulations, we have chosen $\epsilon=0.02,\, \alpha=0.7,\, y_0=0.01,\, B_1= 10^{-3},\, B_2= -2,\, B_3= 3,\, \gamma_\mu^{(0)}= 1,\,\gamma_\nu^{(0)}=2,\,\sigma_ \mu^{(0)}=0,\,\sigma_\nu^{(0)}= 0,\,   \sigma_ \mu^{(1)}=0,\,\sigma_ \mu^{(2)}=0 ,\,\sigma_ \nu^{(1)}=0,\,\sigma_ \nu^{(2)}=0,\,\gamma_\mu^{(1)}= 0,\,\gamma_\mu^{(2)}= 0,\,\gamma_\nu^{(1)}= 0,\,\gamma_\nu^{(2)}= 0$}\label{Fig:caseI-nu}
\end{figure}

The complete solutions up to order $\epsilon^2$ are shown in Figs.~\ref{Fig:caseI-mu} and \ref{Fig:caseI-nu}. As one can see from the differences between the approximate and the exact numerical results obtained from \eqref{Eq:System_KCform}, the method here considered is able to describe the correct behavior of the solution and the absolute error is consistent with the order of the perturbative expansion. 

\subsection{Class II}\label{Sec:case2}

To study these Class II modes we adopt a different method, developed here for the first time, which is apt to describe the fast superimposed oscillations that represent the main effect of bigravity interactions. In this case we have $\alpha \ll \epsilon\ll 1$. Therefore, the pre-factor $\epsilon^2/\alpha^2$ of the last terms in Eqs.~\eqref{Eq:System_KCform_b}, \eqref{Eq:System_KCform_d} is divergent in the limit where both $\alpha$ and $\epsilon$ tend to zero. For this reason, it is convenient to introduce a new quantity $\delta$ defined as
\be
\delta\equiv \left(\frac{\alpha}{\epsilon}\right)^{1/2}\ll1~.
\ee
We also rescale the action and angle variables corresponding to the oscillator $\nu$ as
\be\label{Eq:Def_Iphinu}
{\cal J}_\nu= \delta^{-6} {\cal I}_{\nu}~,\quad \varphi_\nu= \delta^{-2}\phi_\nu ~,
\ee
in order to isolate the divergent behavior controlled by the small parameter $\delta$.
This way, the system \eqref{Eq:System_KCform} can be rewritten as
\begin{subequations}\label{Eq:System_KCform_rescaled}
\begin{align}
\dot{{\cal J}}_\mu&=  \frac{2\epsilon ^2\delta^{-2} q(\tilde{t})\sqrt{{\cal J}_\mu\, {\cal I}_\nu}
   }{\left(r(\tilde{t})+c^2\delta^4\right)^{1/4}\,\left( 1+\epsilon ^2 p(\tilde{t})\right)^{1/4}}\cos(\varphi_\mu) \sin \left(\frac{\phi_\nu}{\delta^{2}}\right)~,\label{Eq:System_KCform_rescaled_a}\\
\dot{{\cal I}}_\nu&=\frac{2s(\tilde{t})  \sqrt{{\cal I}_\nu\,{\cal J}_\mu}}{\left(r(\tilde{t})+c^2\delta^4\right)^{1/4}\, \left(1+\epsilon ^2 p(\tilde{t})\right)^{1/4}}\sin(\varphi_\mu) \cos \left(\frac{\phi_\nu}{\delta^{2}}\right)~,\label{Eq:System_KCform_rescaled_c}\\
\dot{\varphi}_\mu&=  \sqrt{1+\epsilon ^2 p(\tilde{t})} -\frac{ \epsilon ^2\delta^{-2}q(\tilde{t})\sqrt{{\cal I}_\nu } }{\sqrt{{\cal J}_\mu}
   \left(r(\tilde{t})+c^2\delta^4\right)^{1/4}\,\left( 1+\epsilon ^2 p(\tilde{t})\right)^{1/4}} \sin(\varphi_\mu) \sin \left(\frac{\phi_\nu}{\delta^{2}}\right)~,\label{Eq:System_KCform_rescaled_b}\\
\dot{\phi}_\nu&=\sqrt{r(\tilde{t})+c^2\delta^4}-\frac{\delta^2s(\tilde{t}) \sqrt{{\cal J}_\mu }}{\sqrt{{\cal I}_\nu}\left(r(\tilde{t})+c^2\delta^4\right)^{1/4}\, \left(1+\epsilon ^2 p(\tilde{t})\right)^{1/4}} \sin(\varphi_\mu) \sin \left(\frac{\phi_\nu}{\delta^{2}}\right)~.\label{Eq:System_KCform_rescaled_d}
\end{align}
\end{subequations}
Note that, as $\phi_{\nu}$ is real, we must require $r(\tilde{t})>0$ to ensure consistency of the $\delta\to 0$ limit. From the definitions in \eqref{Eq:pqrsDef} we conclude that, since $B_1>0$ and under our assumptions $y_*$ is small and $c>0$ (see \eqref{Eq:LapseApproxSolution_fin}), this condition is satisfied for $\tilde{t}$ up to order 1 as long as $B_1>-(B_2(1+c)+B_3 c y_*)y_*$.

We remark that, upon integration of the highly oscillatory functions in Eq.~\eqref{Eq:System_KCform_rescaled}, one gets a $O(\delta^2)$ contribution since $\int \de t\, F(t)\sin \left(\delta^{-2}\phi(t)\right)\sim O(\delta^2)$ 
(see Appendix~\ref{Sec:AppendixHOQ} for details on the assumptions over the functions $\phi$ and $F$). This is a key property of the system that (upon integration) allows us to keep track of the order of the various terms that arise in the expansion for small $\delta$. In particular, we observe that even though the right-hand sides of Eqs.~\eqref{Eq:System_KCform_rescaled_a} and \eqref{Eq:System_KCform_rescaled_b} involves negative powers of $\delta$, these are cancelled out upon integration, so that there are no divergences in the $\delta\to0$ limit.

Next, we define $A\equiv \sqrt{{\cal I}_\nu}$, $B\equiv \sqrt{{\cal J}_\mu}$ and rewrite our system \eqref{Eq:System_KCform_rescaled} in a simpler form
\begin{subequations}\label{Eq:System_KCform_AB}
\begin{align}
\dot{A}&=\mathcal{F}_{A}(\tilde{t})\,B\sin(\varphi_\mu)\cos\left(\frac{\phi_\nu}{\delta^{2}} \right)~,\label{Eq:System_KCform_AB_a}\\
\dot{B}&=\frac{\epsilon^2}{\delta^{2}}\mathcal{F}_{B}(\tilde{t})\,A\cos(\varphi_\mu)\sin\left(\frac{\phi_\nu}{\delta^{2}} \right)~,\label{Eq:System_KCform_AB_b}\\
\dot{\varphi}_\mu&=  \sqrt{1+\epsilon ^2 p(\tilde{t})} -  \frac{\epsilon ^2}{\delta^{2}} \mathcal{F}_B(\tilde{t})\, \frac{A}{B}
   \sin(\varphi_\mu) \sin \left(\frac{\phi_\nu}{\delta^{2}}\right)~,\label{Eq:System_KCform_AB_c}\\
\dot{\phi}_\nu&=\sqrt{r(\tilde{t})+c^2\delta^4}-\delta^2 \mathcal{F}_A(\tilde{t})\, \frac{B}{A} \sin(\varphi_\mu) \sin \left(\frac{\phi_\nu}{\delta^{2}}\right)~,\label{Eq:System_KCform_AB_d}
\end{align}
\end{subequations}
having defined
\be
\mathcal{F}_{A}(\tilde{t})\equiv s(\tilde{t})\left[\left(r(\tilde{t})+c^2\delta^4\right) \left(1+\epsilon ^2 p(\tilde{t})\right)\right]^{-1/4}~,\quad \mathcal{F}_{B}(\tilde{t})\equiv \frac{q(\tilde{t})}{s(\tilde{t})}\mathcal{F}_{A}(\tilde{t})~.
\ee
We integrate Eqs.~\eqref{Eq:System_KCform_AB_a} and ~\eqref{Eq:System_KCform_AB_b}
\begin{subequations}\label{Eq:System_KCform_ABonly}
\begin{align}
A(t)&=A_0+\int_{t_0}^t \de u\; \mathcal{F}_{A}(\tilde{u})B(u)\sin(\varphi_\mu(u))\cos\left(\frac{\phi_\nu(u)}{\delta^{2}}\right)~,\\
B(t)&=B_0+\frac{\epsilon^2}{\delta^{2}}\int_{t_0}^t \de v\; \mathcal{F}_{B}(\tilde{v})A(v)\cos(\varphi_\mu(v))\sin\left(\frac{\phi_\nu(v)}{\delta^{2}}\right)~,
\end{align}
\end{subequations}
where $u$, $v$ are dummy variables (with $\tilde{u}=\epsilon\, u$, $\tilde{v}=\epsilon\, v$) and $t_0$ is some initial time where we set initial conditions $A_0=A(t_0)$, $B_0=B(t_0)$. The integral equations~\eqref{Eq:System_KCform_ABonly} can be solved iteratively, by repeated substitutions. To include all contributions of order $\epsilon^2$, $\delta^2$, we need two iterations\footnote{We have also computed the triple-integral terms that arise from a further iteration, checking explicitly that they amount indeed to higher-order contributions. Note that we must necessarily expand not only in $\delta$ but also in $\epsilon$ in order to have a finite number of terms. In fact, an expansion just in powers of $\delta$ would lead to infinite iterations and thus infinitely many terms.
}, which gives
\begin{subequations}\label{Eq:System_KCform_ABonly_iterative}
\begin{equation}
\begin{split}
A(t)\approx & A_0+B_0\int_{t_0}^t \de u\; \mathcal{F}_{A}(\tilde{u})\sin(\varphi_\mu(u))\cos\left(\frac{\phi_\nu(u)}{\delta^2}\right)+\\
&\frac{\epsilon^2}{\delta^2} A_0 \int_{t_0}^t \de u \int_{t_0}^u\de v \;  \mathcal{F}_{A}(\tilde{u}) \mathcal{F}_{B}(\tilde{v})\cos(\varphi_\mu(v))\sin\left(\frac{\phi_\nu(v)}{\delta^2}\right)\sin(\varphi_\mu(u))\cos\left(\frac{\phi_\nu(u)}{\delta^2}\right)\,,
\end{split}
\end{equation}
\begin{equation}
\begin{split}
B(t)\approx & B_0+\frac{\epsilon^2}{\delta^2}A_0\int_{t_0}^t \de v\; \mathcal{F}_{B}(\tilde{v})\cos(\varphi_\mu(v))\sin\left(\frac{\phi_\nu(v)}{\delta^2}\right)+\\
&\frac{\epsilon^2}{\delta^2} B_0 \int_{t_0}^t \de v \int_{t_0}^v\de u \;  \mathcal{F}_{B}(\tilde{v})\mathcal{F}_{A}(\tilde{u})\sin(\varphi_\mu(u))\cos\left(\frac{\phi_\nu(u)}{\delta^2}\right) \cos(\varphi_\mu(v))\sin\left(\frac{\phi_\nu(v)}{\delta^2}\right)\,.
\end{split}
\end{equation}
\end{subequations}

Substituting these expansions for $A$ and $B$ into \eqref{Eq:System_KCform_AB_c}, and retaining terms up to second order in $\epsilon$ and $\delta$, we obtain
\be
\dot{\varphi}_\mu\approx  1+\frac{\epsilon ^2}{2} p(\tilde{t}) -  \frac{\epsilon ^2}{\delta^{2}}\frac{A_0}{B_0}q(\tilde{t})\left(r(\tilde{t})\right)^{-1/4}
   \sin(\varphi_\mu) \sin \left(\frac{\phi_\nu}{\delta^{2}}\right)~.\label{Eq:System_KCform_phis_a}
\ee
We proceed similarly for $\phi_\nu$, but in this case we need to include terms up to order $\delta^4$; this is necessary for consistency since $\phi_\nu$ always appears in the combination $\phi_\nu/\delta^2$ in the argument of trigonometric functions. Thus, we obtain
\be
\dot{\phi}_\nu\approx\sqrt{r(\tilde{t})}\left(1+\frac{c^2\delta^4}{2\,r(\tilde{t})}\right)-\delta^2\frac{B_0}{A_0} s(\tilde{t})\left(r(\tilde{t})\right)^{-1/4}
  \sin(\varphi_\mu) \sin \left(\frac{\phi_\nu}{\delta^{2}}\right) ~.\label{Eq:System_KCform_phis_b}
\ee
We can write the solutions to Eqs.~\eqref{Eq:System_KCform_phis_a} and \eqref{Eq:System_KCform_phis_b} as
\begin{subequations}\label{Eq:System_KCform_anglesol}
\begin{align}
\varphi_\mu(t) & \approx  \overline{\varphi}_\mu(t)+\frac{\epsilon^2}{2}\int_{t_0}^t\de u\;p(\tilde{u})-\frac{\epsilon ^2}{\delta^{2}}\frac{A_0}{B_0} \int_{t_0}^{t}\de u\; \frac{q(\tilde{u})}{\left(r(\tilde{u})\right)^{1/4}}
\, 
   \sin(\varphi_\mu(u)) \sin \left(\frac{\phi_\nu(u)}{\delta^{2}}\right)~,\label{Eq:System_KCform_anglesol_a}\\
\phi_{\nu}(t)  & \approx \overline{\phi}_\nu(t)+\frac{c^2\delta^4}{2}\int_{t_0}^t \frac{\de u}{\sqrt{r(\tilde{u})}}-\delta^2\frac{B_0}{A_0} \int_{t_0}^{t}\de u\; \frac{s(\tilde{u})}{\left(r(\tilde{u})\right)^{1/4}}
\, \sin(\varphi_\mu(u)) \sin \left(\frac{\phi_\nu(u)}{\delta^{2}}\right)~,\label{Eq:System_KCform_anglesol_b}
\end{align}
\end{subequations}
having defined
\begin{subequations}\label{Eq:System_KCform_anglesolOverline}
\begin{align}
\overline{\varphi}_\mu(t)&\equiv \varphi_\mu(t_0)+(t-t_0)~,\\
\overline{\phi}_\nu(t)&\equiv\phi_\nu(t_0)+\int_{t_0}^t\de u\; \sqrt{r(\tilde{u})}~.
\end{align}
\end{subequations}
The integrals in \eqref{Eq:System_KCform_anglesol} include higher-order terms that can be removed by iteratively substituting the  right-hand sides of \eqref{Eq:System_KCform_anglesol_a} and \eqref{Eq:System_KCform_anglesol_b} in the argument of the trigonometric functions. At this point, all the arguments in the trigonometric functions are $\overline{\varphi}_\mu$ or $\overline{\phi}_\nu/\delta^2$---both of them being non-oscillatory functions, which allows us to make use of the asymptotic formulae in Appendix~\ref{Sec:AppendixHOQ}. Thus, evaluating the remaining integrals we obtain
\begin{subequations}\label{Eq:System_KCform_angleMuNuSol}
\begin{equation}
\begin{split}
\varphi_\mu(t) & \approx  \overline{\varphi}_\mu(t)+\frac{\epsilon^2}{2}\int_{t_0}^t\de u\;p(\tilde{u})+\epsilon^2\frac{A_0}{B_0}\left[ \frac{q(\tilde{u})}{\left(r(\tilde{u})\right)^{3/4}}\sin(\overline{\varphi}_\mu(u))\cos\left(\frac{\overline{\phi}_\nu(u)}{\delta^2}\right)  \right]_{t_0}^t\\ 
& -\epsilon^2\int_{t_0}^t\de u\; \frac{q(\tilde{u})s(\tilde{u})}{r(\tilde{u})}\sin^2(\overline{\varphi}_\mu(u))~,
\end{split}
\end{equation}
\begin{equation}
\phi_{\nu}(t) \approx \overline{\phi}_\nu(t)+\frac{c^2\delta^4}{2}\int_{t_0}^t \frac{\de u}{\sqrt{r(\tilde{u})}}+\delta^4\left[ \frac{s(\tilde{u})}{\left(r(\tilde{u})\right)^{3/4}}\sin(\overline{\varphi}_\mu(u))\cos\left(\frac{\overline{\phi}_\nu(u)}{\delta^2}\right)  \right]_{t_0}^t~.
\end{equation}
\end{subequations}
Having finally obtained approximate solutions for $\varphi_\mu$ and $\phi_\nu$, $A$ and $B$ can be obtained substituting the expressions \eqref{Eq:System_KCform_angleMuNuSol} in Eq.~\eqref{Eq:System_KCform_ABonly_iterative}, and expanding to order $\epsilon^2$ and $\delta^2$. After integration, the solutions read
\begin{subequations}\label{Eq:System_KCform_AB_sol}
\begin{align}
A(t)&\approx A_0+\delta^2 B_0\left[ \frac{s(\tilde{u})}{\left(r(\tilde{u})\right)^{3/4}}\sin(\overline{\varphi}_\mu(u))\sin\left(\frac{\overline{\phi}_\nu(u)}{\delta^2}\right)  \right]_{t_0}^t~,\\
B(t)&\approx B_0-\epsilon^2 A_0 \left[\frac{q(\tilde{u})}{\left(r(\tilde{u})\right)^{3/4}}\cos(\overline{\varphi}_\mu(u))\cos\left(\frac{\overline{\phi}_\nu(u)}{\delta^2}\right)\right]_{t_0}^t+\frac{\epsilon^2 B_0}{2}\int_{t_0}^t\de u \frac{q(\tilde{u})s(\tilde{u})}{r(\tilde{u})}\sin(2\overline{\varphi}_\mu(u))~.
\end{align}
\end{subequations}

\begin{figure}[hbtp]
 \begin{subfigure}{0.5\textwidth}
        \centering
       \includegraphics[scale=0.18]{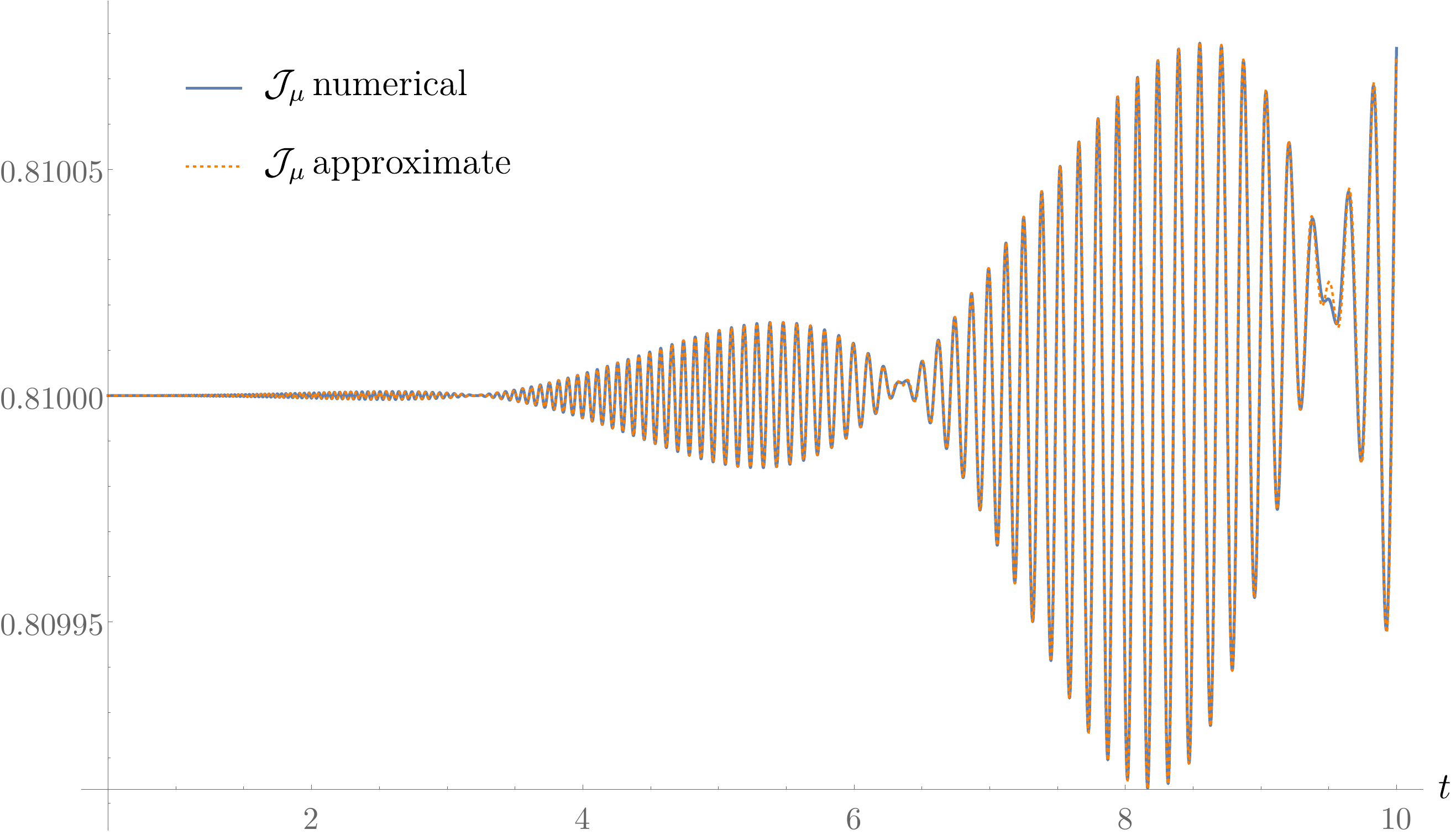}
        \subcaption{}
    \end{subfigure}
     \begin{subfigure}{0.5\textwidth}
        \centering
       \includegraphics[scale=0.173]{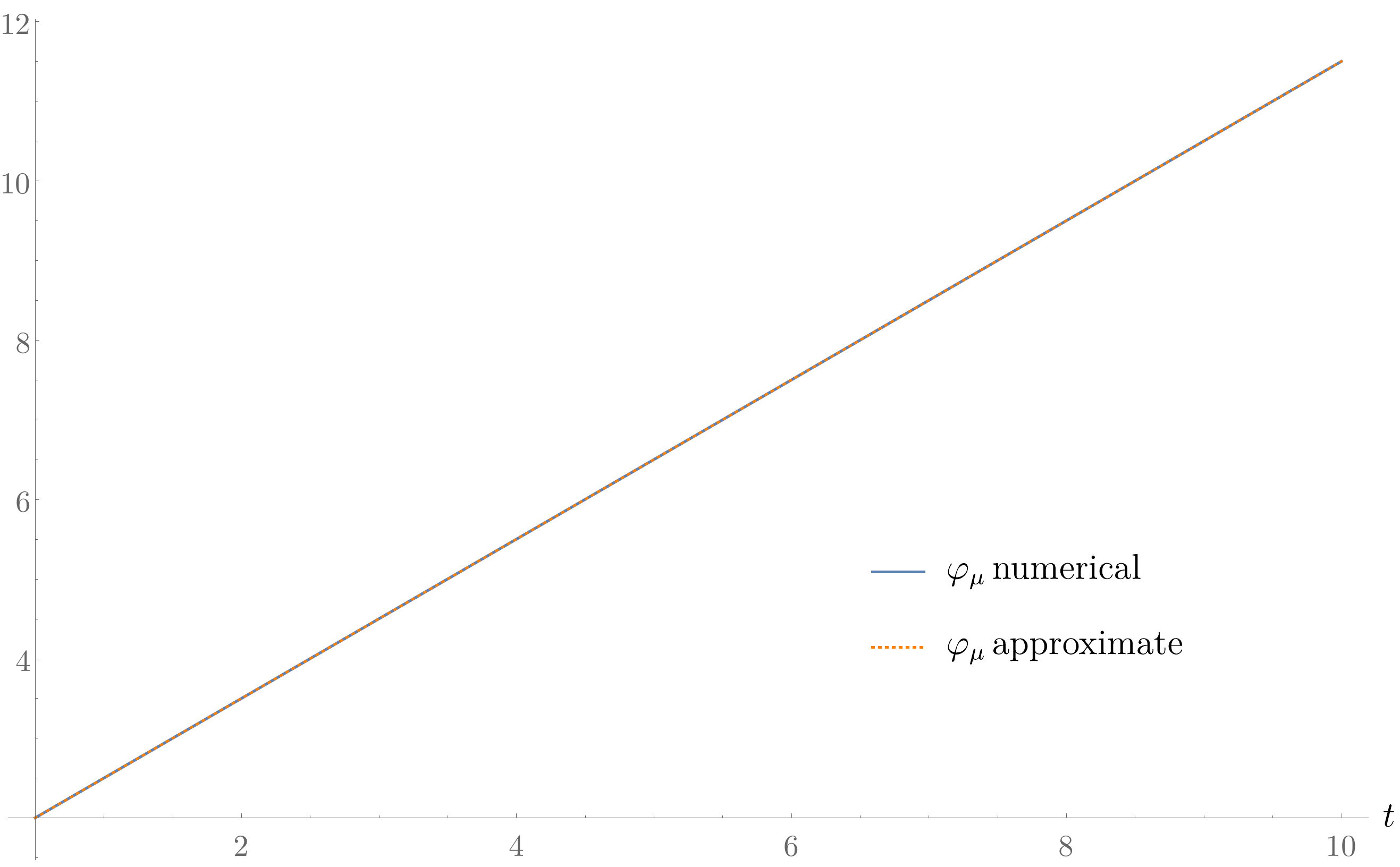}
        \subcaption{}
    \end{subfigure}
     \begin{subfigure}{0.5\textwidth}
        \centering
       \includegraphics[scale=0.18]{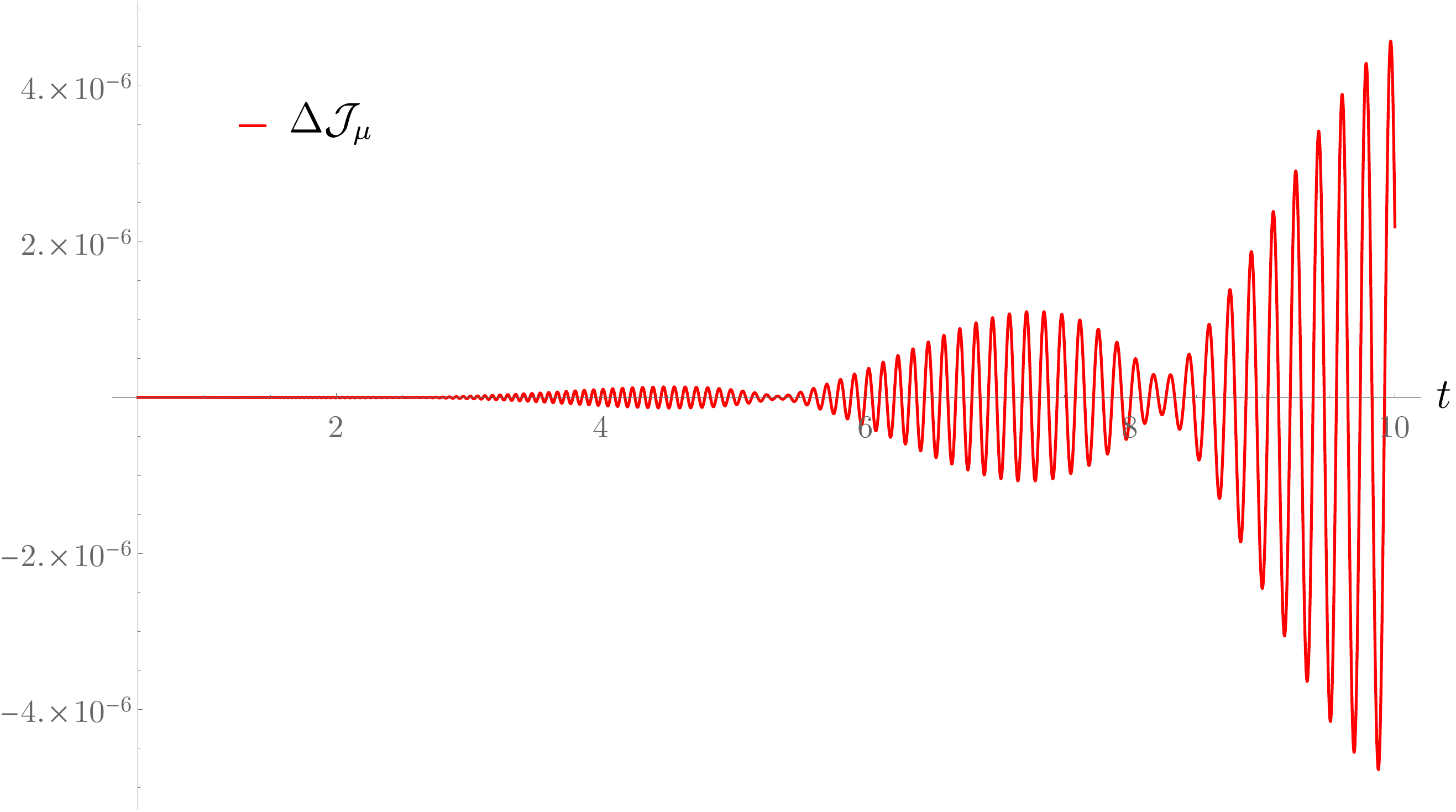}
        \subcaption{}
    \end{subfigure}
    \begin{subfigure}{0.5\textwidth}
        \centering
       \includegraphics[scale=0.18]{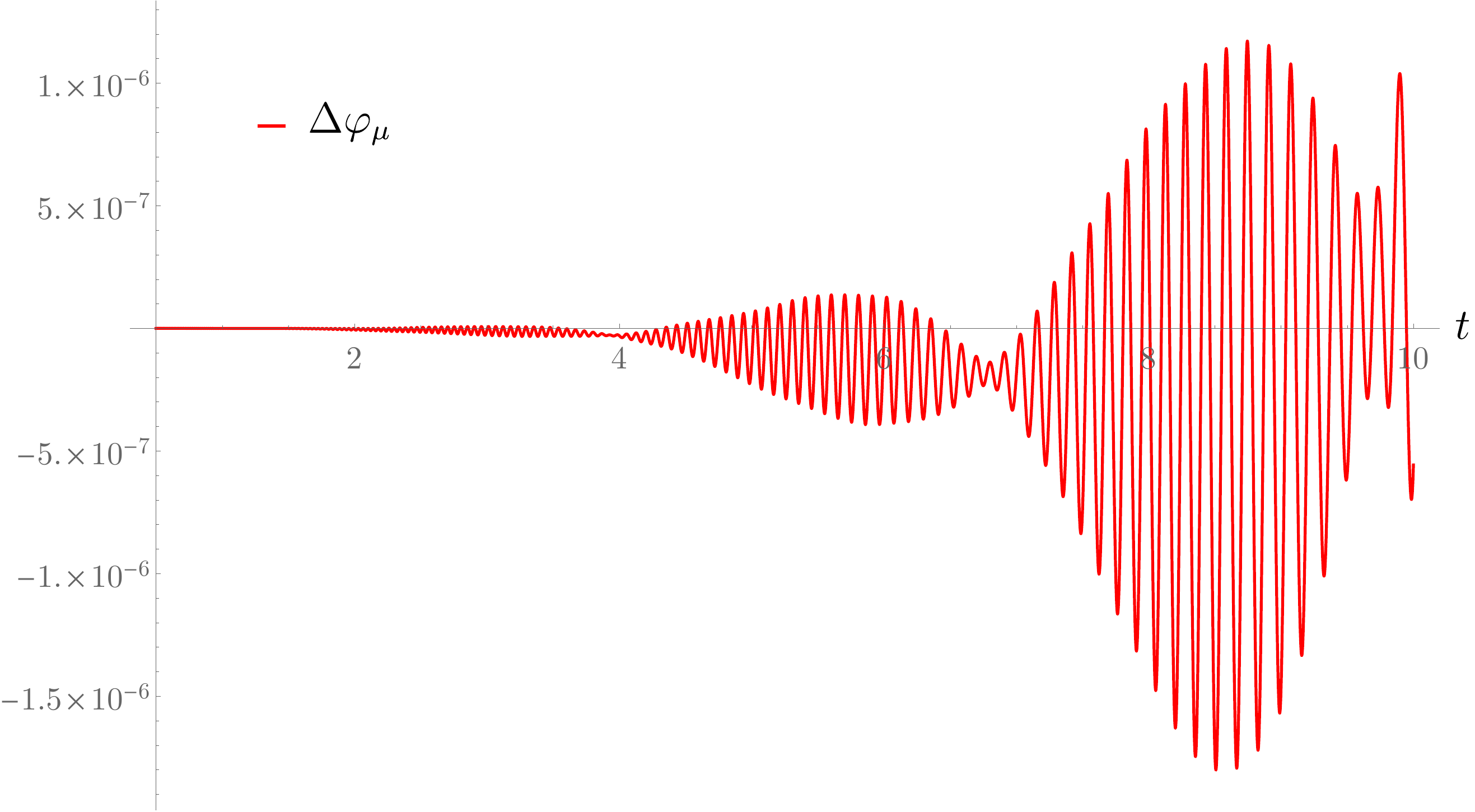}
        \subcaption{}
    \end{subfigure}
\caption{Evolution of the approximate solution obtained in \eqref{Eq:System_KCform_angleMuNuSol} and \eqref{Eq:System_KCform_AB_sol}, and the exact numerical solution of \eqref{Eq:System_KCform} for (a) $\mathcal{J}_{\mu}$ and (b) $\varphi_{\mu}$ for Class II modes during the radiation dominated era. (c) and (d) show the difference between the numerical and the approximate solutions, where $\Delta \mathcal{J}_{\mu}\equiv \mathcal{J}_{\mu \text{\,numerical}}-\mathcal{J}_{\mu \text{\,approximate}}$ and $\Delta \varphi_{\mu}\equiv \varphi_{\mu \text{\,numerical}}-\varphi_{\mu \text{\,approximate}}$. For these simulations, we have chosen $\epsilon=0.1,\,\delta=0.8,\, y_0=0.01,\, B_1= 1,\, B_2= -2,\, B_3= 3,\,t_0=0.5,\,A_0=1,\, B_0=0.9,\,\phi_\nu(t_0)=3/2,\,\varphi_\mu(t_0)= 2$.}\label{Fig:caseII-mu}
\end{figure}

\begin{figure}[hbtp]
 \begin{subfigure}{0.5\textwidth}
        \centering
       \includegraphics[scale=0.18]{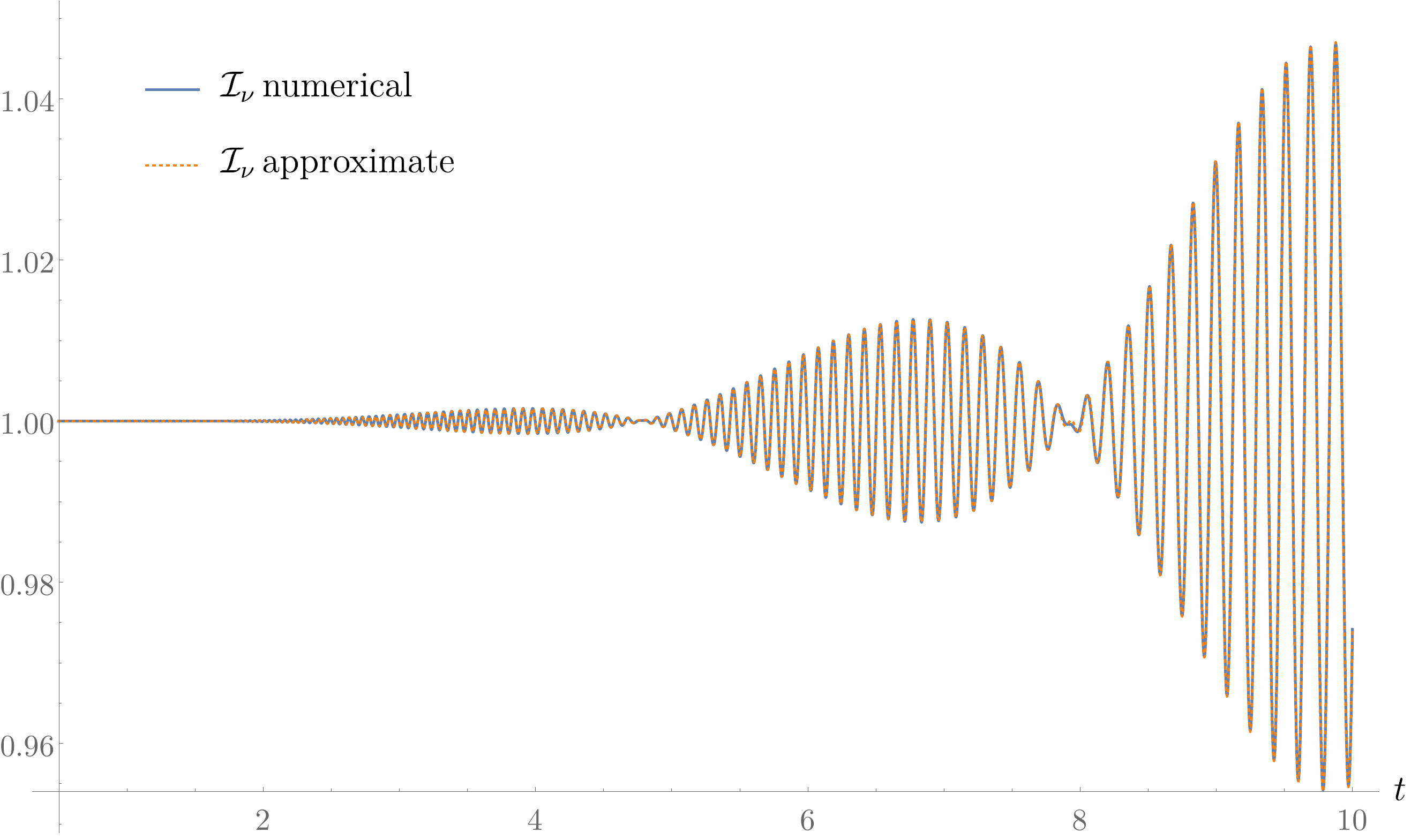}
        \subcaption{}
    \end{subfigure}
     \begin{subfigure}{0.5\textwidth}
        \centering
       \includegraphics[scale=0.173]{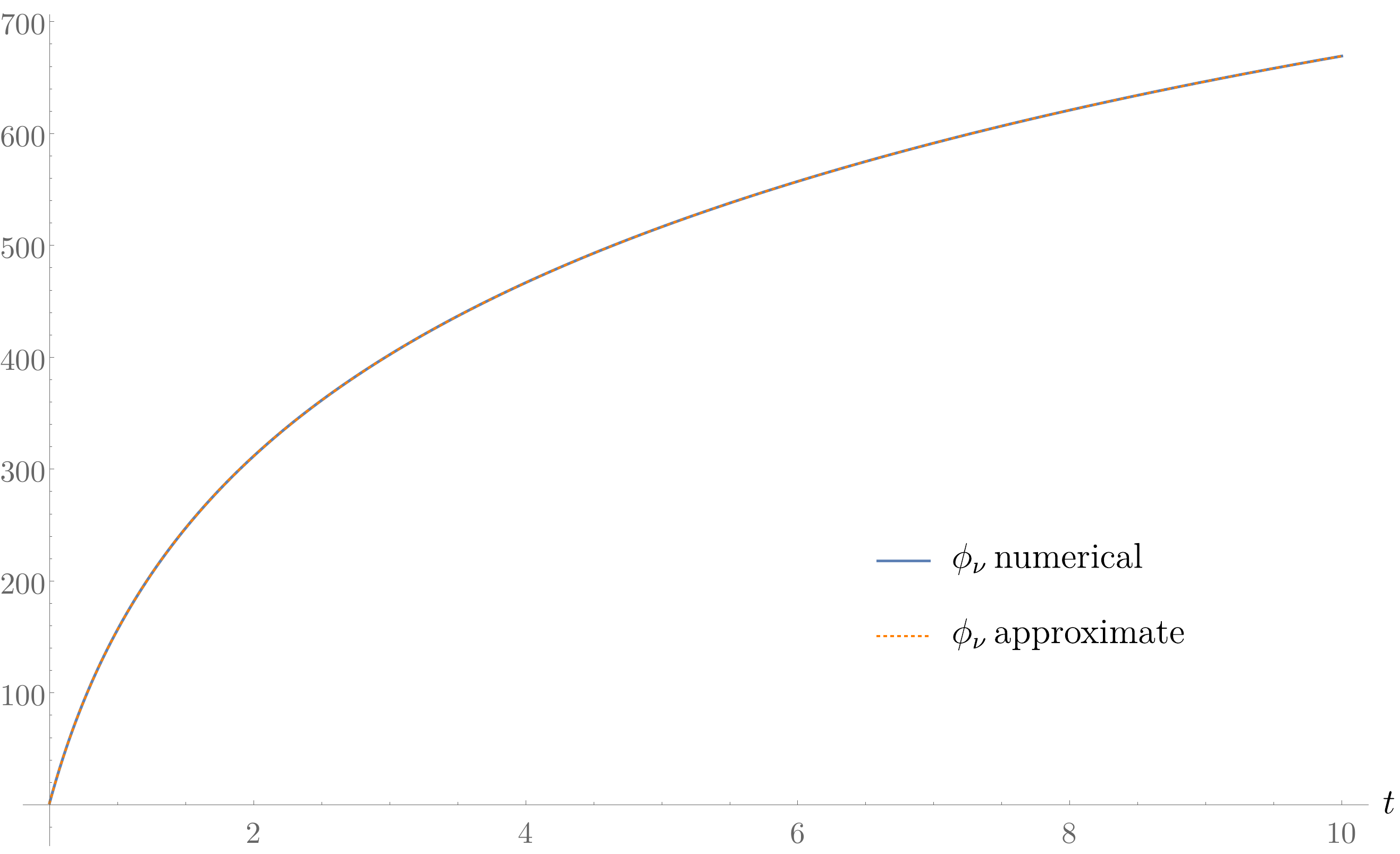}
        \subcaption{}
    \end{subfigure}
     \begin{subfigure}{0.5\textwidth}
        \centering
       \includegraphics[scale=0.18]{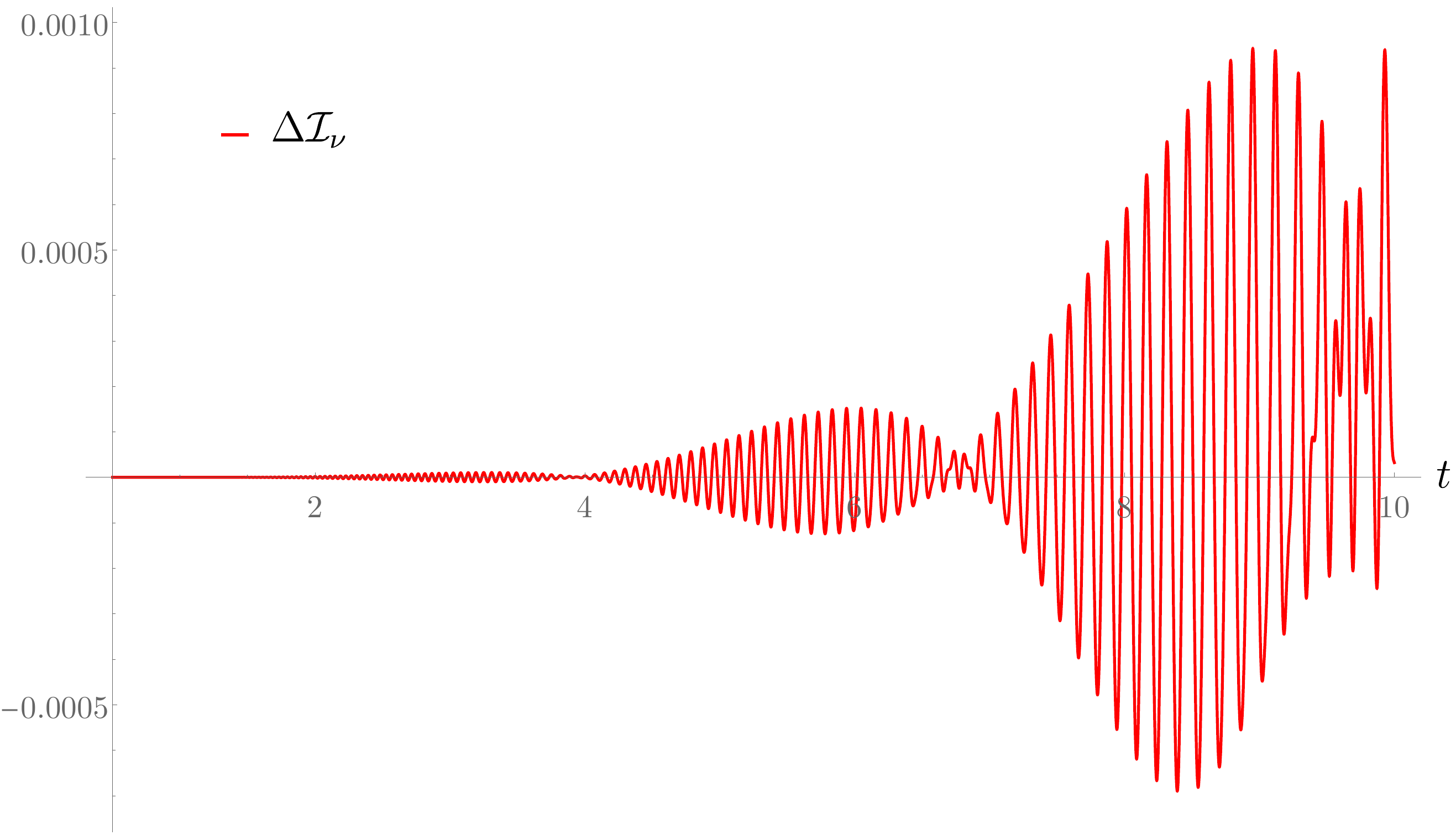}
        \subcaption{}
    \end{subfigure}
    \begin{subfigure}{0.5\textwidth}
        \centering
       \includegraphics[scale=0.18]{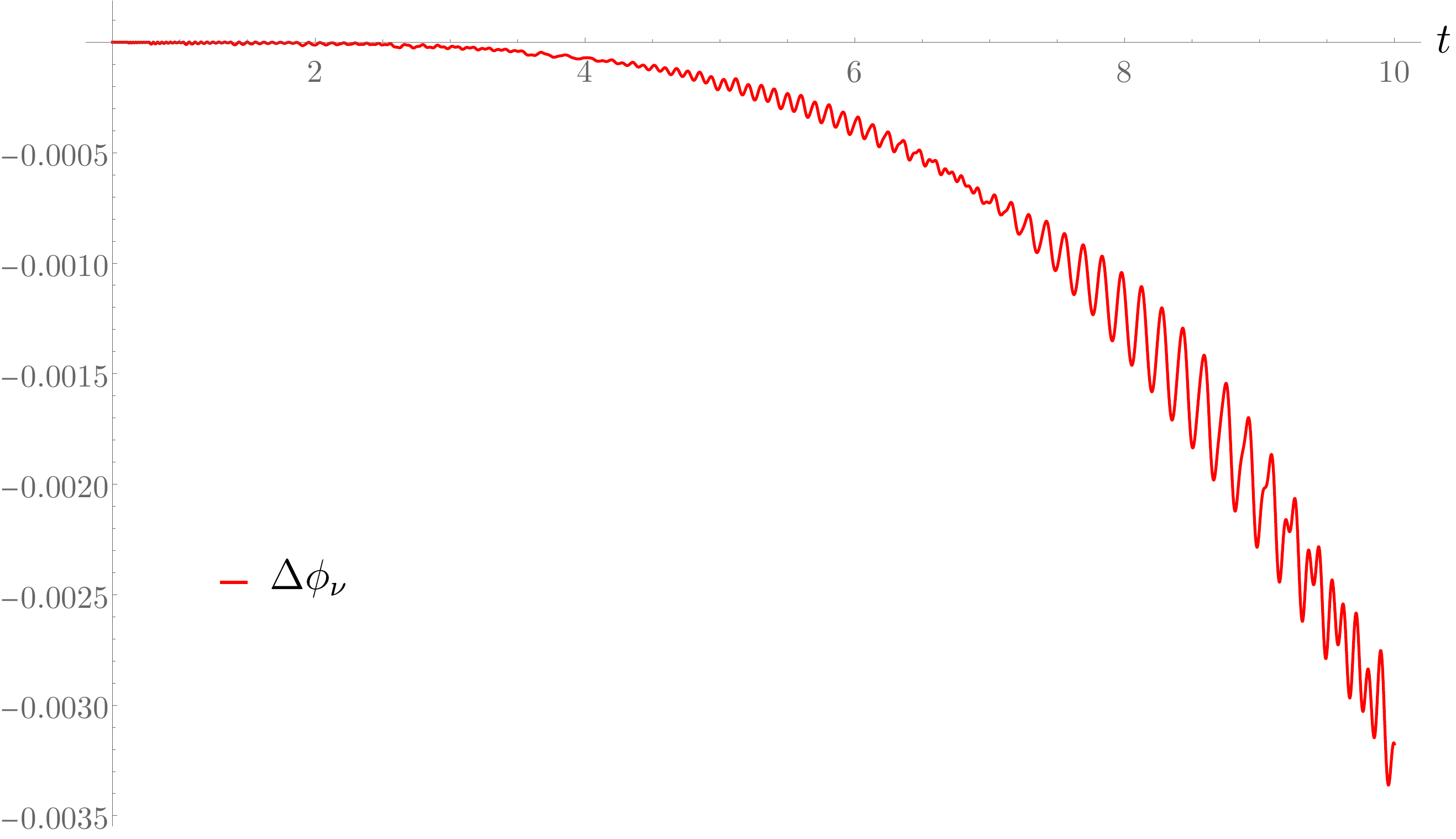}
        \subcaption{}
    \end{subfigure}
\caption{Evolution of the approximate solution obtained in \eqref{Eq:System_KCform_angleMuNuSol} and \eqref{Eq:System_KCform_AB_sol}, and the exact numerical solution of \eqref{Eq:System_KCform} for (a) $\mathcal{I}_{\nu}$ and (b) $\phi_{\nu}$ for Class II modes during the radiation dominated era. (c) and (d) show the difference between the numerical and the approximate solutions, where $\Delta \mathcal{I}_{\nu}\equiv \mathcal{I}_{\nu \text{\,numerical}}-\mathcal{I}_{\nu \text{\,approximate}}$ and $\Delta \phi_{\nu}\equiv \phi_{\nu \text{\,numerical}}-\phi_{\nu \text{\,approximate}}$.  For these simulations, we have chosen $\epsilon=0.1,\,\delta=0.8,\, y_0=0.01,\, B_1= 1,\, B_2= -2,\, B_3= 3,\,t_0=0.5,\,A_0=1,\, B_0=0.9,\,\phi_\nu(t_0)=3/2,\,\varphi_\mu(t_0)= 2$.}\label{Fig:caseII-nu}
\end{figure}

\begin{figure}[hbtp]
 \begin{subfigure}{0.5\textwidth}
        \centering
       \includegraphics[scale=0.18]{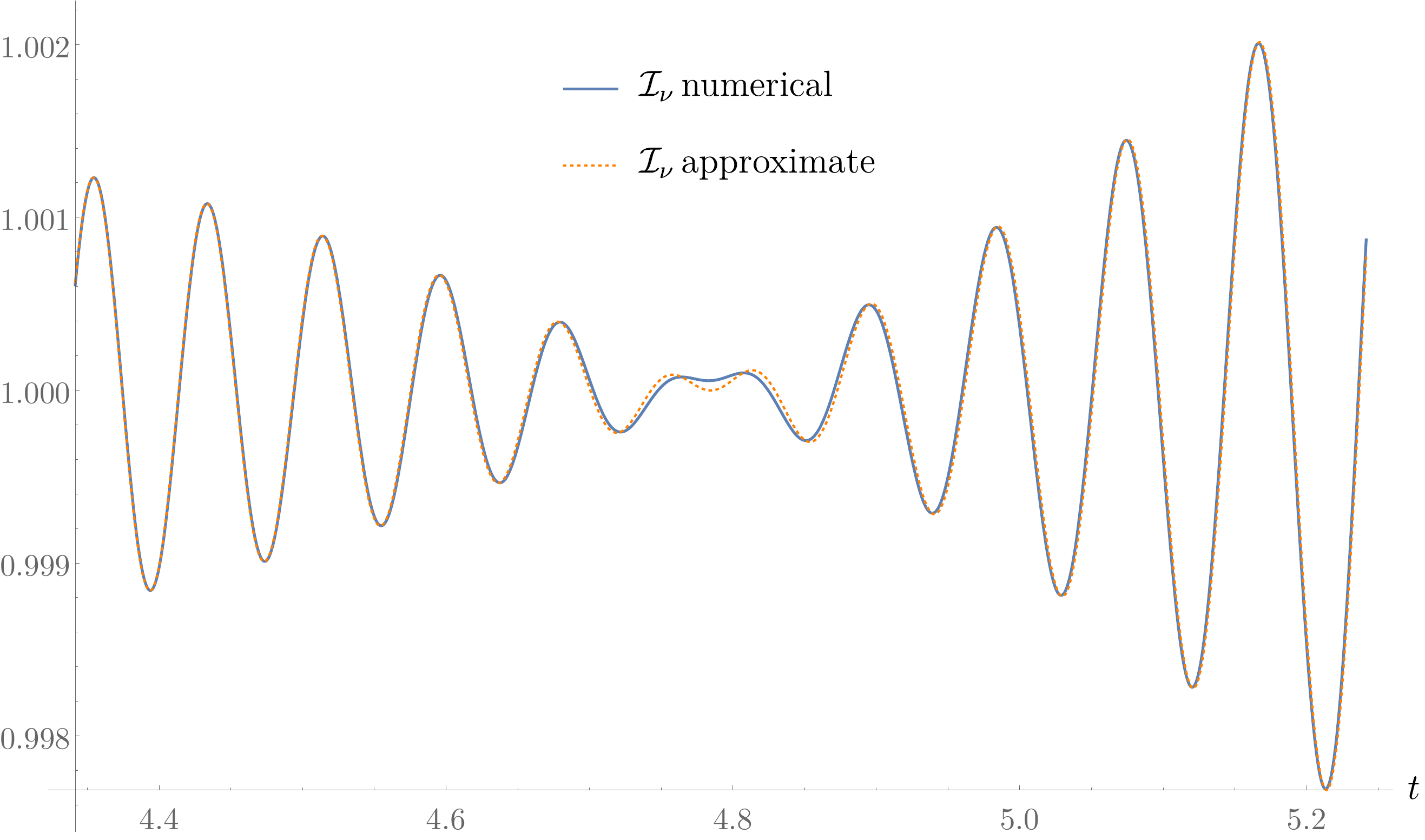}
        \subcaption{}
    \end{subfigure}
     \begin{subfigure}{0.5\textwidth}
        \centering
       \includegraphics[scale=0.173]{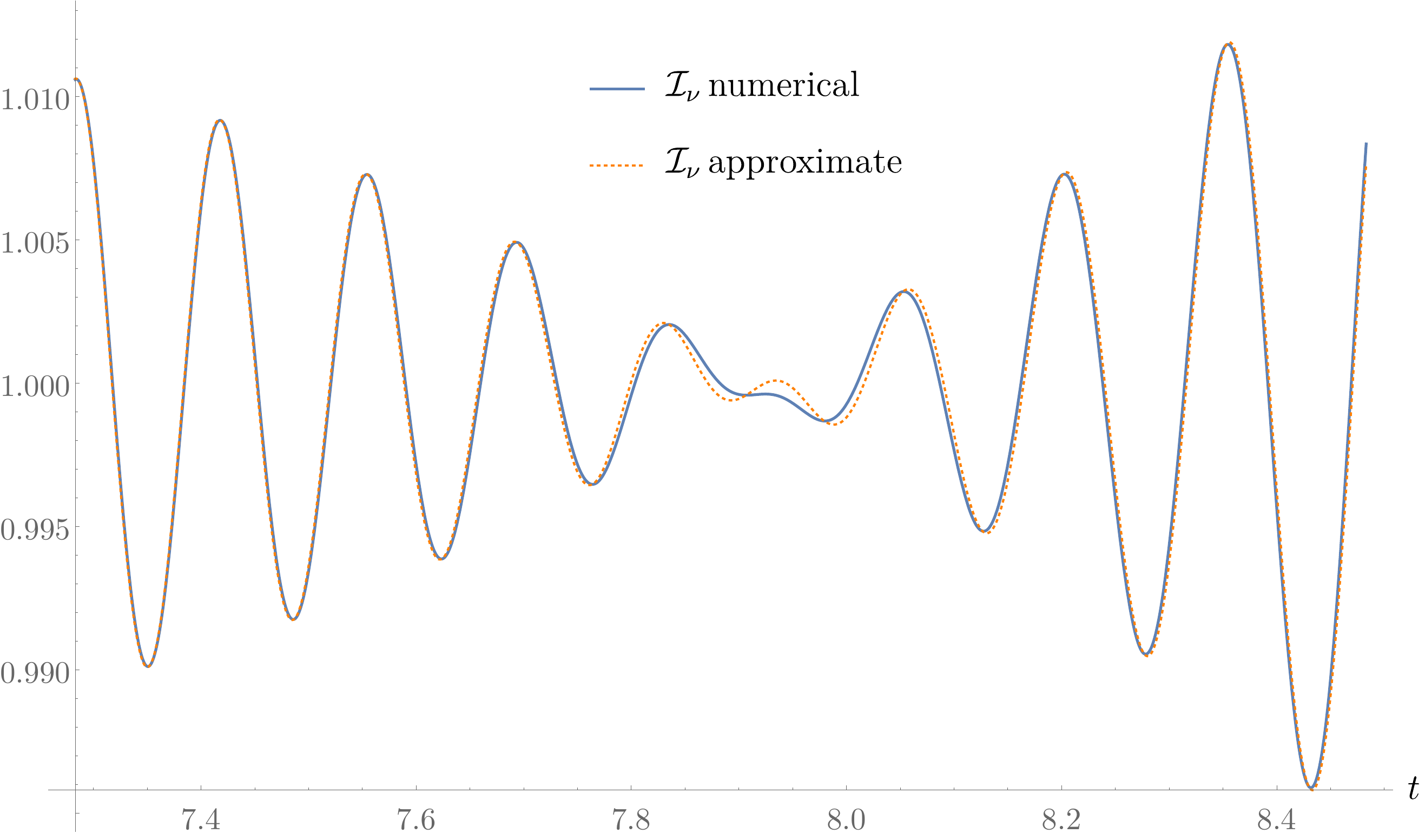}
        \subcaption{}
    \end{subfigure}
\caption{Nodes of the envelope corresponfing to the solution of $\mathcal{I}_{\nu}$ in Fig.\,\ref{Fig:caseII-nu}.}\label{Fig:caseII-nu-nodes}
\end{figure}

Figures \ref{Fig:caseII-mu} and \ref{Fig:caseII-nu} show the evolution of the approximate solutions obtained in \eqref{Eq:System_KCform_angleMuNuSol} and \eqref{Eq:System_KCform_AB_sol}, along with the corresponding numerical solutions of \eqref{Eq:System_KCform}. As in the previous section, the absolute error of the perturbative solutions is consistent with the order of the asymptotic expansion considered. Nevertheless, we note that the approximation here considered breaks down at the nodes of the envelopes of the solution, as it is clear from Fig.\,\ref{Fig:caseII-nu-nodes}. This is because the leading-order asymptotics of the highly oscillatory integrals in \eqref{Eq:System_KCform_AB_sol} vanish at the nodes; therefore, in order to accurately describe this region one should include higher-order terms. More specifically, in our case this means including $O(\delta^4)$ terms in the asymptotic expansions (see Appendix~\ref{Sec:AppendixHOQ}). Thus, since the distance between two nodes is $\Delta t=\pi$, our current solutions describe correctly the superimposed oscillations, although only during a half-period of the $\mu$ modes.

\section{Conclusion}\label{Sec: Conclusion}
In this paper we have analyzed the propagation of cosmological tensor perturbations in bimetric gravity in the GR limit $M_f/M_g\ll 1$, focusing on sub-horizon modes $k\eta\ll1$ in a universe dominated by hydrodynamic matter. Mathematically, the system can be described as a pair of linear oscillators with time-dependent frequencies and couplings. The dynamics is governed by two dimensionless parameters, $\alpha\equiv M_f/M_g$ and $\epsilon\equiv 1/(k\eta_*)$  (where $_*$ denotes the end of the cosmic epoch at hand).
We have identified two main regimes of interest (corresponding to two distinct classes of modes) determined by the relative magnitude of $\epsilon$ and $\alpha$~. There are significant physical differences between such regimes and each requires specific analytical techniques for its study, particularly for the derivation of approximate analytical solutions.

For Class I modes, with $k\gg (M_g/M_f){\cal H}_*$, the couplings of the dynamical system are slowly evolving, and solutions can thus be obtained with multiple-scale analysis, using the methods of Ref.~\cite{kevorkian1996multiple}. In this regime, $\alpha$ can be treated as a finite quantity whereas $\epsilon$ is infinitesimal. After rewriting the system in action-angle variables, a perturbative scheme is adopted where each term of the perturbative series in $\epsilon$ depends on two time variables $t$ and $\tilde{t}\equiv \epsilon t$ (respectively, fast and slow times)---treated as independent. The method enables us to find uniform approximations to the solutions over a time interval of the order $1/\epsilon$ without introducing secular terms. Even though the couplings are slowly evolving, the action variables for the two oscillators are not conserved over time, thus signalling a breakdown of adiabaticity, as shown in Figs \ref{Fig:caseI-mu} and \ref{Fig:caseI-nu}. In terms of the original oscillator variables, we observe that to zeroth and first order in $\epsilon$ the main effect consists of a slow modulation of the amplitude and frequency of the oscillators, while the couplings between oscillators only give rise to higher-order corrections.

On the other hand, for Class II modes (${\cal H}_*\ll k\ll (M_g/M_f){\cal H}_*$) the system is strongly coupled and different analytical techniques are thus required for its study. In this regime both parameters $\alpha$ and $\epsilon$ must be treated as infinitesimal and the evolution of the system is characterized by fast super-imposed oscillations, whose frequency is divergent in the limit where $\delta^2\equiv\alpha/\epsilon$ tends to zero. In order to capture this crucial feature of the system, we have developed a suitable perturbative scheme based on the asymptotics of highly oscillatory integrals, as a double expansion in $\delta$ and $\epsilon$.
Also in this case the action variables are not conserved, and there are significant deviations from adiabaticity due to rapid oscillations on short time scales, as illustrated in Figs.\,\ref{Fig:caseII-mu} and \ref{Fig:caseII-nu}. Also in this case, our approximation scheme is accurate over time scales of the order $t\sim1/\epsilon$. Our approximation could be further improved by systematically including higher-order corrections, which are necessary to properly account for the behavior of the action variables near the nodes of the envelopes, as shown in Fig.~\ref{Fig:caseII-nu-nodes}.

Lastly, we remark that the dynamics of tensor modes in an expanding universe in bimetric gravity has been studied earlier in Ref.~\cite{BeltranJimenez:2019xxx} (see also \cite{LISACosmologyWorkingGroup:2019mwx,Ezquiaga:2021ler}), 
using a scheme that attempts to generalize the standard WKB method to the case of interacting oscillators with time-dependent couplings and friction. The salient features of this approach are reviewed in Appendix~\ref{Sec:AppendixWKB}. Here we show that the method has not been developed in a fully consistent fashion, and provide a concrete example where it fails to provide a reliable approximation to the numerical solution, and it is in fact much less accurate than the multiple-scale method. Thus, future work should be devoted to a careful re-evaluation of the projected constraints on gravitational wave oscillations derived in \cite{LISACosmologyWorkingGroup:2019mwx} using the methods here developed.

\section*{Acknowledgements}
We thank Fernando Barbero for interesting discussions.
The work of MdC is supported by Ministero dell'Universit{\`a} e Ricerca (MUR) (Bando PRIN 2017, 
Codice Progetto: 20179ZF5K5\_006) and by INFN (Iniziative specifiche QUAGRAP and GeoSymQFT).
ASO acknowledges financial support from the fellowship PIF21/237
of the UPV/EHU. This work has been supported by the Basque Government Grant 
\mbox{IT1628-22} and by the Grant PID2021-123226NB-I00 (funded by
MCIN/AEI/10.13039/501100011033 and by ``ERDF A way of making Europe'').

\appendix

\section{Asymptotic formulae for the integration of fast oscillating functions}\label{Sec:AppendixHOQ}

Assuming that $F(t)$ and $\phi(t)$ are nonoscillatory functions and that $\phi(t)$ is free of stationary points---i.e., $\dot{\phi}(t)\neq 0$ within the interval of integration $(t_1,t_2)$--- we have \cite{huybrechs_olver_2009}
\be\label{Eq:HOQ_sin}
\int_{t_1}^{t_2}\de t\, F(t) \sin \left(\frac{\phi(t)}{\delta^2}\right) = -\delta^2 \left[\frac{F(t)}{\dot{\phi}(t)} \cos\left(\frac{\phi(t)}{\delta^2}\right)\right]_{t_1}^{t_2}+\mathcal{O}(\delta^4)~.
\ee
Similarly,
\be\label{Eq:HOQ_cos}
\int_{t_1}^{t_2}\de t\, F(t) \cos \left(\frac{\phi(t)}{\delta^2}\right) = \delta^2\left[\frac{F(t)}{\dot{\phi}(t)} \sin\left(\frac{\phi(t)}{\delta^2}\right)\right]_{t_1}^{t_2}+\mathcal{O}(\delta^4)~.
\ee
These formulae can be extended to the case where $F$ includes trigonometric functions. One just has to combine the oscillatory functions in the integrands using trigonometric identities and then evaluate the resulting integrals using \eqref{Eq:HOQ_cos} and \eqref{Eq:HOQ_sin}. For example,
\begin{equation}
\begin{split}
&\int_{t_0}^t \de u\; \frac{s(\tilde{u})}{\left(r(\tilde{u})\right)^{1/4}}\sin(\overline{\varphi}_\mu(u))\sin\left(\frac{\overline{\phi}_\nu(u)}{\delta^2}\right)\\
&\qquad\qquad\qquad\qquad =-\frac{1}{2}\int_{t_0}^t\de u\; \frac{s(\tilde{u})}{\left(r(\tilde{u})\right)^{1/4}}\left[\cos\left(\frac{\overline{\phi}_\nu(u)}{\delta^2} +\overline{\varphi}_\mu(u)\right)-\cos\left(\frac{\overline{\phi}_\nu(u)}{\delta^2} -\overline{\varphi}_\mu(u)\right)  \right]\\
&\qquad\qquad\qquad\qquad= -\delta^2 \left[ \frac{s(\tilde{u})}{\left(r(\tilde{u})\right)^{1/4}\dot{\overline{\phi}}_\nu(u)}\sin(\overline{\varphi}_\mu(u))\cos\left(\frac{\overline{\phi}_\nu(u)}{\delta^2}\right)  \right]_{t_0}^t+\mathcal{O}(\delta^4)~.
\end{split}
\end{equation}

\section{Second-order action for tensor perturbations}\label{Sec:AppendixSecondOrderAction}

We expand the Hassan-Rosen action~\eqref{Eq:BimetricAction} to second order in the metric perturbations $h_{ij}$, $H_{ij}$, around a solution of the background field equations
\be\label{Eq:ActionExpansion}
S_{\rm\scriptscriptstyle HR}=S_{\rm\scriptscriptstyle HR}^{(0)}+\frac{1}{2!}S_{\rm\scriptscriptstyle HR}^{(2)}+\dots~,
\ee
where $S_{\rm\scriptscriptstyle HR}^{(r)}$ denotes the $r$-th order variation. %, defined following the same conventions as in Ref.~\cite{}.
 Moreover, by definition, the first variation $S_{\rm\scriptscriptstyle HR}^{(1)}$ vanishes if the background obeys the field equations. In this appendix, we will assume the vacuum field equations for reasons of simplicity; however, the generalization to the case where matter is included is straightforward.

In order to evaluate the right-hand side of Eq.~\eqref{Eq:ActionExpansion}, we compute the second order expansion of the matrix $\mathbb{S}$, which appears in the definition of the bigravity interactions,
\be
\mathbb{S} = y \begin{pmatrix} c & 0 \\ 0 & \delta^{i}_{\; j}+\frac{1}{2}(H^{i}_{\;j}-h^{i}_{\;j})+\frac{1}{8}(3 h^{i}_{\;l}+H^{i}_{\;l})(h^{l}_{\;j}-H^{l}_{\;j})+\dots 
  \end{pmatrix} ~.
\ee
Then, recalling the definitions of the symmetric polynomials \eqref{Eq:SymmetricPoly_Defs}, we obtain the following expansions 
\begin{subequations}\label{Eq:SymPolyExpansion}
\begin{align}
e_1 (\mathbb{S})&=y\left[3+c+\frac{1}{8}\left(3 h_{ij}h^{ij} -H_{ij}H^{ij}-2h_{ij}H^{ij}  \right)+\dots \right]  ~,\\
e_2 (\mathbb{S})&=y^2 \left[3(1+c) +\frac{1}{8}\left( (5+3c)h_{ij}h^{ij} -(3+c)H_{ij}H^{ij}-2(1+c) h_{ij}H^{ij}  \right)+\dots \right]  ~,\\
e_3 (\mathbb{S})&=y^3\left[1+3c +\frac{1}{8}\left( (2+5c)h_{ij}h^{ij} -(2+3c)H_{ij}H^{ij}-2 c\, h_{ij}H^{ij}  \right)+\dots \right]~,\\
e_4 (\mathbb{S})&=y^4 c \left[ 1+\frac{1}{4}\left( h_{ij}h^{ij} - H_{ij}H^{ij}\right) +\dots \right]~.
\end{align}
\end{subequations}
For the volume elements we have
\be\label{Eq:VolElsExpansion}
\sqrt{-g}= \sqrt{\gamma}\; a^4 \left(1-\frac{1}{4}(h^{i}_{\;j})^2+\dots \right)~,\quad \sqrt{-f}= \sqrt{\gamma}\; b^4 c \left(1-\frac{1}{4}(H^{i}_{\;j})^2 +\dots \right)~,
\ee
where $\sqrt{\gamma}$ is the spatial volume element associated with $\gamma_{ij}$. Finally, combining Eqs.~\eqref{Eq:SymPolyExpansion} and \eqref{Eq:VolElsExpansion}, and recalling the definitions of the effective pressures \eqref{Eq:EffectiveFluids}, we obtain the following expansion for the interaction terms in the bimetric action \eqref{Eq:BimetricAction},
\be\label{Eq:ZerothOrderInteractions}
\begin{split}
S_{\rm int}^{(0)} & =-m^2 M_g^2 \int \de^4x\sqrt{\gamma}\; a^4 \left( \beta_0 +\beta_1 (3+c) y + 3\beta_2(1+c)y^2+\beta_3(1+3c) y^3+\beta_4 c\, y^4 \right)\\
& = \int \de^4x\sqrt{\gamma} \left( a^4 p_g +b^4 c\, p_f \right)~,
\end{split}
\ee
\be\label{Eq:SecondOrderInteractions}
\begin{split}
S_{\rm int}^{(2)}= \int \de^4x \sqrt{\gamma} &\left[  -\frac{1}{2}\left( a^4 p_g\, h_{ij}h^{ij} +b^4 c\, p_f\, H_{ij}H^{ij}\right) +\right. \\
&\quad \left. -\frac{1}{4}m^2 M_g^2 a^4\left(h_{ij}-H_{ij}\right)\left(h^{ij}-H^{ij}\right)\left(\beta_1 y+\beta_2 (1+c)y^2+\beta_3 c\, y^3\right) \right]~,
\end{split}
\ee

The first two terms in Eq.~\eqref{Eq:BimetricAction} have the same structure as the Einstein-Hilbert action; therefore, their expansion proceeds as in general relativity. We have, up to a total divergence
\be\label{Eq:EHg}
\begin{split}
S_{\scr\rm EH}[g_{ab}]= \frac{M_g^2}{2} \int \de^4 x \sqrt{-g} R^{(g)}=&\frac{M_g^2}{2}  \int \de^4 x \sqrt{\gamma}\; a^2\Bigg[-6\left(\mathcal{H}^2-\frac{1}{\rnew^2}\right)+\frac{1}{4}h^{\prime}_{ij} h^{\prime\;ij}\\
 &- \frac{1}{4}\mathcal{D}_i h_{jk} \mathcal{D}^i h^{jk} -\frac{1}{2}\left(\mathcal{H}^2+2 \mathcal{H}^\prime +\frac{2}{\rnew^2}\right)h_{ij}h^{ij} +\dots\Bigg]~.
\end{split}
\ee
Recalling that the conformal time for $f_{ab}$ satisfies $\de\tilde{\eta}=c(\eta)\de\eta$, from Eq.~\eqref{Eq:EHg} we obtain, after some straightforward substitutions 
\be\label{Eq:EHf}
\begin{split}
S_{\scr\rm EH}[f_{ab}]=\frac{M_f^2}{2}  \int \de^4 x \sqrt{-f} R^{(f)}= & \frac{M_f^2}{2}  \int \de^4 x \sqrt{\gamma}\; b^2 c\Bigg[-6\left(\mathcal{H}_f^2-\frac{1}{\rnew^2}\right)+\frac{1}{4 c^2}H^{\prime}_{ij} H^{\prime\;ij}\\
& - \frac{1}{4}\mathcal{D}_i H_{jk} \mathcal{D}^i H^{jk} -\frac{1}{2}\left(\mathcal{H}_f^2+\frac{2}{c} \mathcal{H}_f^\prime +\frac{2}{\rnew^2}\right)H_{ij}H^{ij}+\dots \Bigg]~.
\end{split}
\ee
Then, combining Eqs.~\eqref{Eq:ZerothOrderInteractions}, \eqref{Eq:EHg} and \eqref{Eq:EHf}, and using the background field equations \eqref{Eq:BackgroundEqs} in vacuo (i.e., with $\rho_m=p_m=0$) we compute the zeroth-order action evaluated on a solution (Hamilton's principal function)
\be
S^{(0)}=-2\int \de^4 x \sqrt{\gamma}\; \left[ M_g^2\, a^2\left(\mathcal{H}^2+2 \mathcal{H}^\prime -\frac{1}{\rnew^2}\right)+M_f^2\, b^2 c \left(\mathcal{H}_f^2+\frac{2}{c} \mathcal{H}_f^\prime -\frac{1}{\rnew^2}\right)\right]~.
\ee
Similarly, from Eqs.~\eqref{Eq:SecondOrderInteractions}, \eqref{Eq:EHg} and \eqref{Eq:EHf}, and using the (vacuum) background field equations \eqref{Eq:BackgroundEqs} to simplify the coefficients of the terms proportional to $h_{ij} h^{ij}$ and $H_{ij} H^{ij}$, we can finally compute the second-order action for tensor perturbations on a curved FLRW background
\be\label{Eq:GWaction}
\begin{split}
 S_{\scr\rm GW}[h_{ij},H_{ij}] &  =\frac{1}{2} \left(S_{\scr\rm EH}^{(2)}[g_{ab}]+S_{\scr\rm EH}^{(2)}[f_{ab}]+S_{\rm int}^{(2)}[g_{ab},f_{ab}]\right)\\
  & =\int \de^4 x \sqrt{\gamma} \Bigg\{ \frac{M_g^2}{8} a^2\left[h^{\prime}_{ij} h^{\prime\;ij} - \mathcal{D}_i h_{jk} \mathcal{D}^i h^{jk} - \frac{2}{\rnew^2}h_{ij}h^{ij}\right] \\
  &+\frac{M_f^2}{8}  b^2 c \left[ \frac{1}{c^2}H^{\prime}_{ij} H^{\prime\;ij} - \mathcal{D}_i H_{jk} \mathcal{D}^i H^{jk}-\frac{2}{\rnew^2} H_{ij}H^{ij} \right]\\
  & -\frac{1}{8}m^2 M_g^2 a^4\left(h_{ij}-H_{ij}\right)\left(h^{ij}-H^{ij}\right)\left(\beta_1 y+\beta_2 (1+c)y^2+\beta_3 c\, y^3\right)   \Bigg\}~.
\end{split}
\ee

The above derivation of the second-order action \eqref{Eq:GWaction} also holds with minimal modifications in the presence of matter fields that contribute to the background field equations. However, in this case there are additional pressure contributions further to the effective pressure terms in \eqref{Eq:SecondOrderInteractions}, leading to analogous cancellations with the background quantities in Eqs.~\eqref{Eq:EHg} and \eqref{Eq:EHf}. Furthermore, even though we focused specifically on a closed FLRW model, the derivation of the second-order action is entirely analogous in open and flat cosmologies. Specifically, the corresponding action for an open universe can be formally obtained from \eqref{Eq:GWaction} with the replacement $\rnew^2 \to - \rnew^2$, while the case of a flat universe is recovered by letting $\rnew^2 \to +\infty$.

Lastly, we observe that in the presence of matter perturbations with a non-vanishing anisotropic stress (i.e., a non-zero transverse traceless component of the perturbed matter stress-energy tensor), here denoted as $\pi_{ij}$, the second-order action includes the extra term
\be\label{Eq:SourceTerm}
S_{\rm source}[h_{ij},\pi_{ij}]=\frac{1}{2}\int \de^4 x \sqrt{\gamma}\; a^4 h^{ij}\pi_{ij}~.
\ee
Here we have assumed that all matter fields only couple to the metric $g_{ab}$, in order to avoid the Boulware-Deser ghost. Therefore, we do not need to introduce an analogous source term for the perturbations of $f_{ab}$.

\section{Shortcomings of the generalized WKB method}\label{Sec:AppendixWKB}

Here we review previous attempts at a generalization of the WKB method, proposed in Refs.~\cite{BeltranJimenez:2019xxx,LISACosmologyWorkingGroup:2019mwx,Ezquiaga:2021ler} to solve similar differential equations as the ones considered in this paper.
A general model is assumed for the cosmological dynamics of coupled tensor modes, which reads (in matrix form)
\be\label{Eq:eom_general}
\left(\hat{I}\frac{\de^2}{\de\eta^2}+\hat{\nu}(\eta)\frac{\de}{\de\eta}+\hat{C}(\eta)k^2+\hat{\Pi}(\eta)k+\hat{M}(\eta) \right) \Phi =0~,
\ee
where 
\be
\Phi=\begin{pmatrix} h \\ H \end{pmatrix}~.
\ee
Here $\hat{I}$ is the identity matrix, $k$ is the spatial momentum, and $\hat{\nu}$, $\hat{C}$, $\hat{\Pi}$, $\hat{M}$ are the friction, velocity, chirality and mass matrices, respectively.  The bigravity case, Eq.~\eqref{Eq:SystemTensorPerthH}, can be obtained as a special case of Eq.~\eqref{Eq:eom_general}. In analogy with the standard WKB method \cite{Bender_1999}, a small dimensionless parameter~$\delta$ (in this case unrelated to the physical parameters of the model) is then introduced by hand to suppress time derivatives
\be\label{Eq:eom_general1}
\left(\delta^2\,\hat{I}\frac{\de^2}{\de\eta^2}+\epsilon\,\hat{\nu}(\eta)\frac{\de}{\de\eta}+\hat{C}(\eta)k^2+\hat{\Pi}(\eta)k+\hat{M}(\eta) \right) \Phi=0~.
\ee
In this way, the determination of $\Phi$ is then framed as finding an asymptotic solution to the singular perturbation problem \eqref{Eq:eom_general1} in the $\delta\to0$ limit. The following ansatz is then made, as a generalization of the usual WKB ansatz that also accounts for mixing
\be\label{Eq:wkb_ansatz}
\Phi=\hat{E}\, e^{\frac{i}{\delta} \int d\eta\,\hat{\theta} }\left(\Phi_0+\delta\,\Phi_1+\dots \right)~,
\ee
where $\hat{E}$ is the matrix of eigenvectors and $\hat{\theta}$ is the diagonal matrix of eigenfrequencies.\footnote{Hereafter, we omit the $\eta$-dependence to make the notation lighter.}

The ansatz \eqref{Eq:wkb_ansatz} is then substituted in the equations of motion~\eqref{Eq:eom_general1}, which gives
\be\label{Eq:wkb_expansion}
{\cal E}_0+{\cal E}_1\,\delta+{\cal E}_2\,\delta^2+\dots=0~,
\ee
where the first few ${\cal E}_n$ are \cite{BeltranJimenez:2019xxx}
\begin{subequations}\label{Eq:wkb_ords012}
\begin{align}
{\cal E}_0&=\left[-\hat{E}\hat{\theta}^2+i\hat{\nu}\hat{E}\hat{\theta}+\left(\hat{C}k^2+\hat{\Pi}k+\hat{M}\right)\hat{E}\right]\hat{G}\hat{\Phi}_0~,\label{Eq:wkb_0}\\
{\cal E}_1&=(2\hat{E}\hat{\theta}-i\hat{\nu}\hat{E})\hat{G}\Phi_0^{\prime}+(\hat{E}\hat{\theta}^{\prime}+2\hat{E}^{\prime}\hat{\theta}-i\hat{\nu}\hat{E}^{\prime})\hat{G}\Phi_0~,\\
{\cal E}_2&=\hat{E}\hat{G}\Phi_0^{\prime\prime}+2\hat{E}^{\prime}\hat{G}\Phi_0^{\prime}+\hat{E}^{\prime\prime}\hat{G}\Phi_0+i(2\hat{E}\hat{\theta}-i\hat{\nu}\hat{E})\hat{G}\Phi_1^{\prime}+i(\hat{E}\hat{\theta}^{\prime}+2\hat{E}^{\prime}\hat{\theta}-i\hat{\nu}\hat{E}^{\prime})\hat{G}\Phi_1~.
\end{align}
\end{subequations}
Here $\hat{G}\equiv e^{\frac{i}{\delta} \int d\eta\, \hat{\theta}}$~, which has an {\it essential singularity} at $\delta=0$~.
Clearly, if the ${\cal E}_n$ were independent of $\delta$~, Eq.~\eqref{Eq:wkb_expansion} would then imply that ${\cal E}_n=0$ for all $n$~. However, this is not the case in general, due to the presence of the matrix $\hat{G}$, which introduces a non-analytic dependence on $\delta$ that cannot be factored out---except for the trivial case where the matrix $\hat{\theta}$ is a multiple of the identity. For this reason, one cannot conclude from \eqref{Eq:wkb_expansion} that ${\cal E}_n=0$~.
However, if one {\it imposes} ${\cal E}_n=0$ as in Refs.~\cite{BeltranJimenez:2019xxx}, correction terms beyond the leading-order approximation in Eq.~\eqref{Eq:wkb_ansatz} unavoidably introduce an exponential sensitivity on $\delta$~.  In Ref.~\cite{BeltranJimenez:2019xxx}, this is masked in part by the fact that the authors set $\delta=1$ after deriving Eqs.~\eqref{Eq:wkb_ords012}.

\begin{figure}[t]
 \begin{subfigure}{1\textwidth}
        \centering
       \includegraphics[scale=0.35]{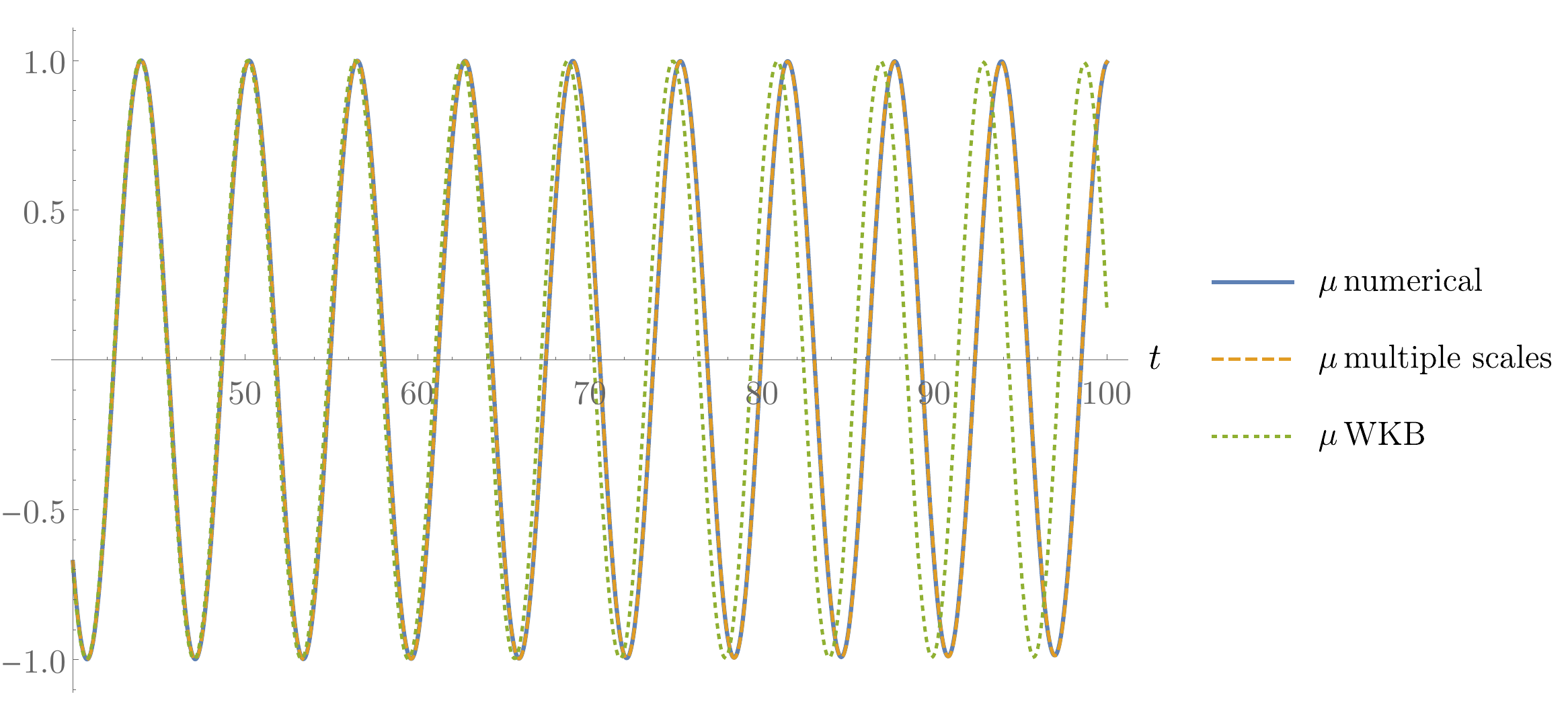}
        \subcaption{}
    \end{subfigure}
     \begin{subfigure}{1\textwidth}
        \centering
       \includegraphics[scale=0.35]{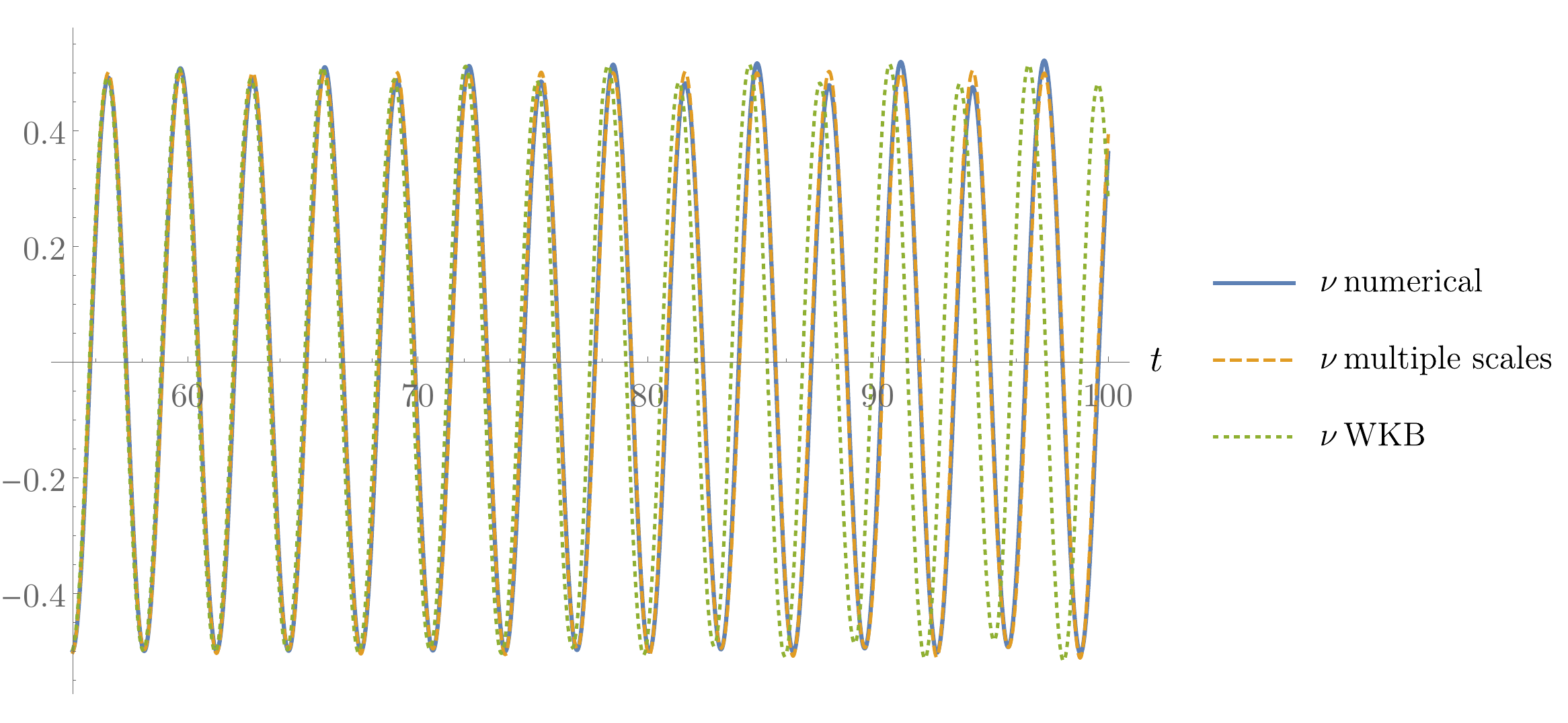}
        \subcaption{}
    \end{subfigure}
\caption{Comparison of the numerical solution of the system~\eqref{Eq:eom_general} particularized for \eqref{Eq: WKBCounterexample} with the multiple-scale and WKB approximations. The parameters of the model are $\epsilon=0.01,\,p=1,\,q=1,\,r=2,\,k=1$ . The integration constants have been fixed at $t=0$, setting $\mu(0)=1,\, \mu^{\prime}(0)=0, \,\nu(0)=0.5,\, \nu^{\prime}(0)=0$.}\label{Fig:WKB}
\end{figure}

The zeroth order matrix equation \eqref{Eq:wkb_0} can be solved to determine $\hat{\theta}$ and $\hat{E}$, see Sec.~2.1.1 of Ref.~\cite{BeltranJimenez:2019xxx} for details. Next, to first order in $\delta$ one obtains
\be\label{Eq:wkb1}
(2\hat{E}\hat{\theta}-i\hat{\nu}\hat{E})\hat{G}\Phi_0^{\prime}+(\hat{E}\hat{\theta}^{\prime}+2\hat{E}^{\prime}\hat{\theta}-i\hat{\nu}\hat{E}^{\prime})\hat{G}\Phi_0=0~.
\ee
In Ref.~\cite{BeltranJimenez:2019xxx} it is noted that Eq.~\eqref{Eq:wkb1} does not admit a closed-form analytical solution and the following expression has been proposed as an approximate solution
\be\label{Eq:phi0_wkbsol}
\Phi_0=\hat{\theta}^{-1/2} e^{-\int d\eta\, \hat{A}_{\scriptscriptstyle\rm WKB}}C_0~,
\ee
where $C_0$ is a constant vector fixed by the initial conditions, and
\be
 \hat{A}_{\scriptscriptstyle\rm WKB}= \hat{G}^{-1}\hat{\theta}^{1/2}\left(2\hat{E}\hat{\theta}-i\hat{\nu}\hat{E}\right)^{-1}\left(2\hat{E}^{\prime}\hat{\theta}-i\hat{\nu}\hat{E}^{\prime}+\frac{i}{2}\hat{\nu}\hat{E}\hat{\theta}^{\prime}\hat{\theta}^{-1}\right)\hat{G}\hat{\theta}^{-1/2}~.
\ee
In Fig.~\ref{Fig:WKB} we compare the numerical solution of the system~\eqref{Eq:eom_general} with the corresponding multiple-scale and WKB approximations, in a toy model where
\be\label{Eq: WKBCounterexample}
\hat{\nu}=\hat{\Pi}=0~,\quad \hat{C}= \begin{pmatrix} 1 & 0 \\ 0 & 4 \end{pmatrix} ~,\quad \hat{M}= \epsilon \begin{pmatrix} a(\tilde{t}) & b(\tilde{t}) \\ w(\tilde{t})a(\tilde{t}) & w(\tilde{t})b(\tilde{t}) \end{pmatrix}~,
\ee 
with $a(\tilde{t})=p\, \tilde{t}^{\,3}+q\, \tilde{t}^{\,4}$, $b(\tilde{t})=\tilde{t}^{\,2} a(\tilde{t})$~, $w(\tilde{t})=r\, \tilde{t}^{-1}$~. The plot clearly shows that the generalized WKB method fails to give an accurate approximation of the numerical solution, which is instead well approximated by the multiple-scale solution.

\bibliographystyle{bib-style}
\bibliography{references_bimetric}

\end{document}